\documentclass[paper,toc,nohyper]{JHEP3}
\pdfoutput=1

\usepackage{ams} 
\usepackage{graphicx}
\usepackage{amsmath}
\usepackage{amsbsy}
\usepackage{booktabs}
\usepackage{pifont}
\usepackage{mathtools}
\usepackage{cite}

\def\b#1{\bar{#1}}
\def\h#1{\hat{#1}}
\def\t#1{\tilde{#1}}
\def\wt#1{\widetilde{#1}}
\def\Poles{{\cal P}oles}
\def\dsigma{{\rm d} \hat\sigma}

\def\ba{\begin{eqnarray}}
\def\ea{\end{eqnarray}}
\def\la{\langle}
\def\ra{\rangle}
\def\bs{\boldsymbol} 
\def\wt{\widetilde}
\def\t{\tilde}
\def\bt{\bs{T}}
\def\e{\epsilon}
\def\nn{\nonumber}
\def\NF{N_F}
\def\b#1{\bar{#1}}
\def\bb#1{\bar{\bar{#1}}}
\def\hb#1{\hat{\bar{#1}}}
\def\hbb#1{\hat{\bar{\bar{#1}}}}
\def\S{{\cal{S}}}

\def\calF{{\cal F}_{3}^{0}}
\def\JET{J}

\allowdisplaybreaks

\preprint{
  IPPP/13/87 \\
  LPN13-073  \\
  ZU-TH 24/13
  \\
  \today}

\title{NNLO QCD corrections to jet production at hadron colliders from gluon scattering}

\author{James Currie$^{a}$, Aude Gehrmann-De Ridder$^{b}$, E.W.N. Glover$^{b,c}$, Jo\~{a}o Pires$^{b}$\\
$^a$Institut f\"ur Theoretische Physik, Universit\"at Z\"urich, Wintherturerstrasse 190,\\
CH-8057, Z\"urich, Switzerland\\
$^b$Institute for Theoretical Physics, ETH, CH-8093 Z\"urich, Switzerland\\
$^c$Institute for Particle Physics Phenomenology, University of Durham,
South Road,\\ Durham DH1 3LE, England}

\abstract{We present the next-to-next-to-leading order (NNLO) QCD corrections to dijet production in the purely gluonic channel retaining the full dependence on the number of colours. The sub-leading colour contribution in this channel first appears at NNLO and, as expected, increases the NNLO correction by around 10\% and exhibits a $p_{T}$ dependence, rising from
8\% at low $p_{T}$ to 15\% at high $p_{T}$. The present calculation demonstrates the utility of the antenna subtraction method for computing the full colour NNLO corrections to dijet production at the Large Hadron Collider.
\\
\today
}

\begin{document}

\section{Introduction}

The dominant hard scattering process at the LHC is the production of 
hadronic jets.  Their ubiquity in the collider environment allows for 
high precision studies of jets to be performed by the 
ATLAS~\cite{atlasjet,atlasjet2} and CMS~\cite{cmsjet,cmsjet2,cmsjet3,cmsjet4} experiments. Jet cross 
sections are theoretically interesting as they are sensitive to the value of the 
strong coupling 
constant~\cite{Giele:1995kb,CDF:2001hn,asjet,atlasrunning}, as well as 
the parton distribution functions~\cite{Giele:1994xd} and to new 
physics beyond the Standard Model~\cite{ATLAS:2012pu,CMS:2013qha}.

In order to fully utilise these precisely measured observables, we need 
a comparably precise understanding of the theoretical prediction for the 
cross section.  The jet cross section can be calculated using a 
combination of perturbative techniques for the hard scattering 
subprocesses and non-perturbative parton distribution functions,
\ba
{\rm d}\sigma&=&\sum_{a,b}\int\frac{{\rm d}\xi_{1}}{\xi_{1}}\frac{{\rm 
d}\xi_{2}}{\xi_{2}}~f_{a}(\xi_{1},\mu^2)f_{b}(\xi_{2},\mu^2)~\dsigma_{ab}(\alpha(\mu^2),\mu^2),
\ea
where the sum runs over parton species in the colliding hadrons. The 
parton distribution function (PDF), 
$f_{a}(\xi_{1},\mu^2){\rm{d}\xi_{1}}$ describes the probability to find 
the parton of species $a$ with momentum fraction $\xi_{1}$ in the 
hadron, which is defined by the choice of factorization scale, $\mu$, 
which in this paper is set equal to the renormalization scale. The 
partonic cross section, $\dsigma_{ab}$, describes the probability for 
the initial-state partons to interact and produce a final-state, $X$, 
normalised to the hadron-hadron flux.

The partonic cross section is calculable within perturbative QCD and has 
a series expansion in the strong coupling constant,
\ba
\dsigma_{ab}&=&\dsigma_{ab,LO}+\bigg(\frac{\alpha_{s}(\mu^2)}{2\pi}\bigg)\dsigma_{ab,NLO}+\bigg(\frac{\alpha_{s}(\mu^2)}{2\pi}\bigg)^2\dsigma_{ab,NNLO}+{\cal{O}}(\alpha_{s}(\mu^2)^3),
\ea
where the series has been truncated at next-to-next-to leading order 
(NNLO). For dijet production the leading order cross section carries an 
overall factor of $\alpha_{s}^2$ such that the NLO and NNLO 
corrections carry overall factors of $\alpha_{s}^{3}$ and 
$\alpha_{s}^{4}$  respectively. The most accurate theoretical 
predictions for dijet observables are currently those calculated at NLO 
accuracy~\cite{Ellis:1988hv,Ellis:1990ek,eks,jetrad,nlojet1,nlojet2,powheg2j,meks}. 
Further improvements include the inclusion of LO~\cite{Baur:1989qt} and NLO~\cite{Moretti:2006ea,Dittmaier:2012kx} 
electroweak corrections and the study of QCD threshold corrections~\cite{Kidonakis:2000gi,Kumar:2013hia}.

As the LHC experiments continue to record and analyse jet data, the 
experimental precision on the single inclusive and exclusive dijet cross 
sections demand better precision from our theory predictions. This has 
led to a drive to provide the NNLO corrections to the jet cross section 
in order to bring theory uncertainties in line with the experimental 
precision attainable at the LHC. Many techniques have been developed in 
recent years to calculate NNLO corrections with hadronic initial 
states.  The antenna subtraction method~\cite{GehrmannDeRidder:2005cm} 
was developed for $e^{+}e^{-}$ annihilation, where it was successfully 
applied to the calculation of the three-jet cross section at 
NNLO~\cite{our3j1,our3j2,ourevent1,ourevent2,ourevent3,weinzierl3j1,weinzierl3j2,weinzierlevent1,weinzierlevent2}. 
The method has subsequently been generalised to hadronic 
initial-states~\cite{Daleo:2006xa,Daleo:2009yj,Boughezal:2010mc,GehrmannDeRidder:2012ja,Gehrmann:2011wi}, 
and applied to the leading colour contributions to gluonic dijet 
production in gluon 
fusion~\cite{Glover:2010im,GehrmannDeRidder:2011aa,GehrmannDeRidder:2012dg,GehrmannDeRidder:2013mf} 
and quark-antiquark annihilation~\cite{Currie:2013vh}. Recent years have 
also seen the development of the sector improved subtraction technique, 
STRIPPER~\cite{Czakon:2010td} which has subsequently been applied to several 
phenomenological studies for top pair 
production~\cite{Czakon:2011ve,czakontop1,czakontop2,czakontop3,czakontop4} 
and Higgs plus jet production~\cite{Boughezal:2013uia}.
The NNLO corrections for a wide range of processes involving the 
production of colourless particles are also known, either for single 
particle production, 
Higgs~\cite{babishiggs,babishiggs1,babishiggs2,babishiggs3, 
grazzinihiggs}, 
Drell-Yan~\cite{babisdy1,babisdy2,kirilldy1,kirilldy2,grazzinidy1,grazzinidy2}, 
and di-boson production 
~\cite{grazziniwh,babisgg,grazzinigg,grazzinizg,deFlorian:2013jea}.

The NNLO mass factorised partonic cross section is composed of three 
contributions: the double real, real-virtual and double virtual corrections,
\ba
\dsigma_{ab,NNLO}&=&\int_{n+2}\dsigma_{ab,NNLO}^{RR}+\int_{n+1}\Big[\dsigma_{ab,NNLO}^{RV}+\dsigma_{ab,NNLO}^{MF,1}\Big]\nn\\
&+&\int_{n}\Big[\dsigma_{ab,NNLO}^{VV}+\dsigma_{ab,NNLO}^{MF,2}\Big],
\ea
where each contribution is defined to contain the relevant phase space 
integration measure and so $\int_{n}$ simply keeps track of the number 
of final-state particles involved in the phase space integral. It is 
well known that each of these terms is separately divergent, either 
containing singularities in regions of single or double unresolved phase 
space or explicit IR poles in $\e$, yet the sum of all three 
contributions can be arranged such that all singularities cancel to 
yield a finite result.  In order to perform this reorganisation we 
construct three subtraction terms such that the partonic cross section 
can be re-expressed in the form,
\ba
\dsigma_{ab,NNLO}&=&\int_{n+2}\Big[\dsigma_{ab,NNLO}^{RR}-\dsigma_{ab,NNLO}^{S}\Big]\nn\\
&+&\int_{n+1}\Big[\dsigma_{ab,NNLO}^{RV}-\dsigma_{ab,NNLO}^{T}\Big]\nn\\
&+&\int_{n\phantom{+1}}\Big[\dsigma_{ab,NNLO}^{VV}-\dsigma_{ab,NNLO}^{U}\Big].
\ea
The double real subtraction term is constructed to remove all single and 
double unresolved divergences and renders the double real channel IR 
finite. The real-virtual subtraction term is a combination of double 
real subtraction terms integrated over a single unresolved phase space, 
mass factorization contributions and new subtraction terms introduced to remove 
the remaining singularities of the real-virtual contribution.  The double 
virtual subtraction term is constructed from mass factorization terms 
and the remaining subtraction terms from the double real and 
real-virtual, integrated over the double and single unresolved phase 
spaces respectively,
\ba
\dsigma_{ab,NNLO}^{T}&=&\dsigma_{ab,NNLO}^{V,S}-\dsigma_{ab,NNLO}^{MF,1}-\int_{1}\dsigma_{ab,NNLO}^{S},\\
\dsigma_{ab,NNLO}^{U}&=&-\dsigma_{ab,NNLO}^{MF,2}-\int_{1}\dsigma_{ab,NNLO}^{V,S}-\int_{2}\dsigma_{ab,NNLO}^{S}.
\ea

In this paper we are concerned with the NNLO correction to the dijet 
cross section in the all-gluon approximation. To help organise the 
calculation it is useful to define the operators ${\cal{LC}}$ and 
${\cal{SLC}}$, which project out the leading colour and sub-leading 
colour corrections such that,
\ba
\dsigma_{gg,NNLO}&=&{\cal{LC}}\Big(\dsigma_{gg,NNLO}\Big)+{\cal{SLC}}\Big(\dsigma_{gg,NNLO}\Big),
\ea
where ${\cal{LC}}\Big(\dsigma_{gg,NNLO}\Big)$ was discussed in Refs.~ 
\cite{Glover:2010im,GehrmannDeRidder:2011aa,GehrmannDeRidder:2012dg,GehrmannDeRidder:2013mf} 
while ${\cal{SLC}}\Big(\dsigma_{gg,NNLO}\Big)$ constitutes the remaining 
contribution to the cross section discussed in this 
paper.\footnote{Note that the definition of the leading colour 
contribution contains an overall factor of $(N^2 -1)$, as does the 
subleading colour contribution.  The two are separated by a relative 
factor of $N^2$ and so strictly expanding as a series in $N$ leads to a 
mixing of the two contributions.  In this paper we \emph{define} the 
${\cal{LC}}$ and ${\cal{SLC}}$ operators to both contain this overall 
factor of $(N^{2}-1)$ so as to avoid such mixing of terms.}

It can be seen by simple power counting in $N$ that the NNLO mass 
factorization terms for this process only contribute to the leading 
colour cross section, i.e.,
\ba
{\cal{SLC}}\Big(\dsigma_{gg,NNLO}^{MF,1}\Big)&=&0,\label{eq:slcmf1}\\
{\cal{SLC}}\Big(\dsigma_{gg,NNLO}^{MF,2}\Big)&=&0\label{eq:slcmf2}.
\ea
The significance of Eqs.~\eqref{eq:slcmf1} and~\eqref{eq:slcmf2} for 
this calculation is that there is no mass factorization contribution at sub-leading colour.

The sub-leading colour contribution poses an interesting theoretical 
challenge for the antenna subtraction scheme previously employed to 
compute the leading colour 
contribution~\cite{Glover:2010im,GehrmannDeRidder:2011aa,GehrmannDeRidder:2012dg,GehrmannDeRidder:2013mf}. 
This method is well suited to leading colour 
calculations and those where the cross section can be written as a sum 
of colour ordered squared partial amplitudes with simple factorization 
behaviour in unresolved 
limits. 
However, the sub-leading colour contribution, is constructed from the 
incoherent interference of partial amplitudes and it is an interesting 
question to see whether the method is sufficiently general to 
systematically remove all of the IR singularities.
As we will show, this can be achieved in a straightforward manner 
without the need to derive new antennae or to perform new analytic 
integrals.

The phenomenology of this process is also interesting as it gives a 
concrete example of the size of sub-leading colour corrections to the 
leading colour process at NNLO. Na\"ively we expect sub-leading colour 
contributions to be numerically small because in the all-gluon channel 
they are suppressed by a factor of $1/N^2$ relative to the leading 
colour contribution. In addition to this power counting, QCD displays 
colour coherence and so sub-leading colour contributions can contain 
incoherent interferences of partial amplitudes.  These incoherent 
interferences will generically contain contributions which are 
suppressed by quantum mechanical destructive interference effects, and 
so the colour incoherent sub-leading colour contributions may be 
suppressed even further than the na\"ive $1/N^2$ suppression. These 
heuristic arguments are appealing but it is also desirable to make firm 
quantitative statements about the relevance of sub-leading colour 
contributions.  In this paper we do so by explicitly calculating the 
sub-leading colour contribution to dijet production at NNLO in the 
all-gluon approximation and comparing it with the leading colour contribution.

The paper is organised in the following way. In Section 2, we define the 
notation used throughout the paper and
introduce also the notions of colour space that help organise the 
sub-leading colour contributions.
In Sections 3, 4 and 5 we systematically step through the  double real, 
real-virtual and double virtual contributions,
first defining the relevant matrix element and then deriving the 
appropriate subtraction terms.   We show that the antenna subtraction 
technique requires no significant alterations or new ingredients in 
order to deal with the incoherent interferences of partial amplitudes. 
In particular, in Section 3 we show that the single and double 
unresolved limits of the double real matrix element at sub-leading 
colour can be fully described using just three-parton tree-level 
antennae, without the need for four-parton antenna functions. In Section 
4, we give a more compact form for the real-virtual matrix element than 
that present in the literature~\cite{Bern:1993mq}. As in the double unresolved case, we 
show that the single unresolved limits of the real-virtual matrix 
element do not require the one-loop three-parton antenna and can be 
described with only tree-level three-parton antennae to remove all 
explicit and implicit singularities. We derive the 
double virtual subtraction term by integrating the remaining double real 
and real-virtual subtraction terms, and show that it analytically 
cancels the explicit poles in the formula for the two-loop 
matrix elements~\cite{Glover:2001af,Glover:2001rd}. We have implemented 
these terms into a parton-level event generator, which can compute the 
all-gluon contribution to any infrared-safe observable related to dijet 
final states at hadron colliders. Section 5 is devoted to a first 
numerical study of the size of the full colour NNLO cross section for 
some experimentally relevant observables; the single jet inclusive distribution for a 
range of rapidity intervals and the dijet invariant mass distribution. 
Finally, our findings are briefly summarized in Section 6.

\section{Notation and colour space}

Throughout this paper, complex amplitudes are denoted by calligraphic letters, whereas real squared amplitudes, summed over helicities are denoted by Roman letters. Generic amplitudes, independent of the scattering process, are written using the letter ${\cal M}$, whereas for the specific process of gluon scattering the we use the letter ${\cal A}$.  Amplitudes and squared amplitudes containing colour information are written in boldface whereas colour stripped amplitudes are not.  Thus, the full $n$-point $\ell$-loop amplitude is denoted by $\bs{\cal M}_{n}^{\ell}$, whereas the same quantity for gluon scattering is denoted by $\bs{\cal A}_{n}^{\ell}$. The corresponding colour stripped partial amplitudes and their squares are denoted by ${\cal M}_{n}^{\ell}$, ${\cal{A}}_{n}^{\ell}$ and $M_{n}^{\ell}$, $A_{n}^{\ell}$ respectively.  The squared full amplitudes, containing all colour information, for generic and gluonic scattering process are denoted by $\bs{M}_{n}^{\ell}$ and $\bs{A}_{n}^{\ell}$.

Specific combinations of integrated antennae and mass factorisation kernels can be used to express the explicit IR poles of one- and two-loop contributions to the cross section.  This approach is of particular use in the antenna subtraction process where writing the poles of the virtual and double virtual cross sections in terms of integrated dipoles allows the pole cancellation to be carried out in a transparent fashion.

The poles of the integrated dipoles correspond to those of the one- and two-loop insertion operators, and so they can be dressed with colour charge operators and inserted into the matrix element sandwiches to obtain the pole structure of the cross section by working in colour space. For $n$-parton scattering, the amplitudes carry colour indices $\{c\}=\{c_{1},\cdots,c_{n}\}$ where $c_{i}=1,\cdots,N^2-1$ for gluons and $c_{i}=1,\cdots,N$ for quarks and antiquarks. A set of basis vectors for the colour space can be constructed, $\{|\bs{c}\ra\}=\{|c_{1}\cdots c_{n}\ra\}$, the projection of an arbitrary vector into which defines a scalar in colour space.  In this space we define a vector which represents a scattering process, such that its projection onto the colour basis vectors produces the coloured scattering amplitude,
\ba
\bs{{\cal M}}_{n}^{\{c\}}(\{p\})&=&\la \bs{c}|{\cal M}_{n}(\{p\})\ra.
\ea
The full squared amplitude, summed over colours is then given by,
\ba
\bs{M}_{n}(\{p\})&=&\sum_{\{c\}}\la{\cal M}_{n}(\{p\})|\bs{c}\ra\la \bs{c}|{\cal M}_{n}(\{p\})\ra\nn\\
&=&\la{\cal M}_{n}(\{p\})|{\cal M}_{n}(\{p\})\ra.
\ea
The emission of a gluon from parton $i$ is associated with the colour charge operator, $\bt_{i}=T_{i}^{c}|c\ra$, which carries the vector colour index of the emitted gluon, $c$, and is a matrix in the colour indices of the emitting parton $i$, i.e.,
\ba
\la \bs{a}|T_{i}^{c}|\bs{b}\ra&=&\delta_{a_{1}b_{1}}\cdots T_{a_{i}b_{i}}^{c}\cdots\delta_{a_{n}b_{n}}.
\ea
The colour charges form an algebra, the elements of which satisfy the following properties,
\ba
\bt_{i}\cdot\bt_{j}&=&\bt_{j}\cdot\bt_{i},\nn\\
\bt_{i}^2&=&C_{i}\bs{1},
\ea
where $\bt_{i}\cdot\bt_{j}=\sum_{c}T_{i}^{c}T_{j}^{c}$ and $\bs{1}$ is the identity matrix in colour space. $C_{i}$ is the Casimir coefficient associated with a parton of type $i$, i.e., for partons in the fundamental representation, $C_{q}=C_{\b{q}}=C_{F}=\frac{N^2-1}{2N}$, for  partons in the adjoint representation, $C_{g}=C_{A}=N$. The product of two colour charges, $\bt_{i}\cdot\bt_{j}$, is a matrix acting on the colour indices of the partons $i$ and $j$ in the scattering process and so when sandwiched between two state vectors, produces a scalar in colour space called a colour correlated matrix element,
\ba
\la{\cal{M}}_{n}|\bt_{i}\cdot\bt_{j}|{\cal{M}}_{n}\ra&=&T_{a_{i}b_{i}}^{c}T_{a_{j}b_{j}}^{c}\bs{\cal{M}}_{n}^{\dagger,\{a\}}(\{p\})\bs{\cal{M}}_{n}^{\{b\}}(\{p\}).\label{eq:mtijm}
\ea
At NNLO we also encounter the colour correlated double operator insertion sandwich, defined by,
\ba
\la{\cal{M}}_{n}|(\bt_{i}\cdot\bt_{j})(\bt_{k}\cdot\bt_{l})|{\cal{M}}_{n}\ra&=&T_{a_{i}b_{i}}^{c_1}T_{a_{j}b_{j}}^{c_1}T_{a_{k}b_{l}}^{c_2}T_{a_{k}b_{l}}^{c_2}\bs{\cal{M}}_{n}^{\dagger,\{a\}}(\{p\})\bs{\cal{M}}_{n}^{\{b\}}(\{p\}).\label{eq:mtijtklm}
\ea
To write down the pole structure of the one- and two-loop cross sections encountered in this paper we must evaluate the following colour charge sandwiches,
\ba
&&\la{\cal{A}}_{n}^{0}|\bt_{i}\cdot\bt_{j}|{\cal{A}}_{n}^{0}\ra,\ n=4,5,\label{eq:titj}\\
&&\la{\cal{A}}_{4}^{0}|\bt_{i}\cdot\bt_{j}|{\cal{A}}_{4}^{1}\ra,\label{eq:titjm1}\\
&&\la{\cal{A}}_{4}^{0}|(\bt_{i}\cdot\bt_{j})(\bt_{k}\cdot\bt_{l})|{\cal{A}}_{4}^{0}\ra.\label{eq:titjtktl}
\ea
For gluons the explicit form of the colour charge operators is given by,
\ba
T_{ab}^{c}&=&if_{acb}.\label{eq:tfabc}
\ea
We choose to write the amplitudes in a colour ordered basis in terms of colour ordered partial amplitudes.  In such a basis, the tree-amplitudes have the form,
\ba
\bs{\cal{A}}_{n}^{0,\{a\}}(\{p\})&=&\sum_{\sigma\in S_{n}/Z_{n}}{\rm{Tr}}(a_{\sigma(1)},\cdots,a_{\sigma(n)})\ {\cal{A}}_{n}^{0}(\sigma(1),\cdots,\sigma(n)),\label{eq:treegluon}
\ea
where the symmetry group $S_{n}/Z_{n}$ contains all non-cyclic permutations of $n$ elements and the arguments of the colour stripped partial amplitudes represent external momenta.  Each $a_{i}$ in the trace of Eq.~\eqref{eq:treegluon} represents a generator of the SU($N$) algebra in the fundamental representation carrying the adjoint colour index $a_{i}$ associated with gluon $i$.  The four-gluon one-loop amplitude, in a colour ordered basis, is given by~\cite{Bern:1990ux},
\ba
\hspace{-0.5cm}\bs{\cal{A}}_{4}^{1,\{a\}}(\{p\})&=&\sum_{\sigma\in S_{4}/Z_4}N\ {\rm{Tr}}(a_{\sigma(1)},a_{\sigma(2)},a_{\sigma(3)},a_{\sigma(4)})\ {\cal{A}}_{4,1}^{1}(\sigma(1),\sigma(2),\sigma(3),\sigma(4))\nn\\
&+&\sum_{\rho\in S_{4}/Z_{2}\times Z_{2}}{\rm{Tr}}(a_{\rho(1)}a_{\rho(2)}){\rm{Tr}}(a_{\rho(3)},a_{\rho(4)})\ {\cal{A}}_{4,3}^{1}(\rho(1),\rho(2),\rho(3),\rho(4)),
\ea
where $\sigma$ is the set of orderings inequivalent under cyclic permutations and $\rho$ is the set of orderings inequivalent under cyclic permutations of the two subsets of orderings $\{\rho(1),\rho(2)\}$ and $\{\rho(3),\rho(4)\}$ and the interchange of these sub-sets. The colour indices of these amplitudes are then contracted with those of the colour charge operators, given in Eq.~\eqref{eq:tfabc}, and conjugate amplitudes to produce the sandwich, as shown in Eqs.~\eqref{eq:mtijm} and~\eqref{eq:mtijtklm}.

A result which will prove useful throughout this paper is that the four-parton tree-level single insertion sandwich in Eq.~\eqref{eq:titj} only contributes at leading colour such that,
\ba
{\cal{SLC}}\Big(\la{\cal{A}}_{4}^{0}|\bt_{i}\cdot\bt_{j}|{\cal{A}}_{4}^{0}\ra\Big)&=&0.\label{eq:slczero}
\ea
Setting $\NF=0$ (according to the all gluon approximation of this paper), the single unresolved integrated dipoles~\cite{Currie:2013vh}, $\bs{J}_{2}^{(1)}$, which dress these colour charges are defined as combinations of integrated antenna functions~\cite{GehrmannDeRidder:2005cm,Daleo:2006xa,Daleo:2009yj,Boughezal:2010mc,GehrmannDeRidder:2012ja,Gehrmann:2011wi} and mass factorisation kernels.  The final-final, initial-final and initial-initial gluon-gluon dipoles are given by,
\ba
\bs{J}_{2}^{(1)}(1_{g},2_{g})&=&\frac{1}{3}{\cal{F}}_{3}^{0}(s_{{1}{2}}),\label{eq:j21def1}\\
\bs{J}_{2}^{(1)}(\hb{1}_{g},2_{g})&=&\frac{1}{2}{\cal{F}}_{3,g}^{0}(s_{\b{1}{2}})-\frac{1}{2}{\Gamma}_{gg}^{(1)}(x_{1})\delta(1-x_{2}),\label{eq:j21def2}\\
\bs{J}_{2}^{(1)}(\hb{1}_{g},\hb{2}_{g})&=&{\cal{F}}_{3,gg}^{0}(s_{\b{1}\b{2}})-\frac{1}{2}{\Gamma}_{gg;gg}^{(1)}(x_{1},x_{2})\label{eq:j21def3},
\ea
where hatted arguments denote initial-state partons and the mass factorization kernels used to define the initial-final and initial-initial dipoles are defined as~\cite{GehrmannDeRidder:2011aa},
\ba
\Gamma_{gg;gg}^{(1)}(x_{1},x_{2})&=&\Gamma_{gg}^{(1)}(x_{1})\delta(1-x_{2})+\Gamma_{gg}^{(1)}(x_{2})\delta(1-x_{1}),\\
\Gamma_{gg}^{(1)}(x_{1})&=&\frac{1}{\e}~p_{gg}^{0}(x_{1}).
\ea
These integrated dipoles can be stitched together to form an integrated antenna string which contains the poles of an extended string of gluons including, by definition, a correlation between the endpoints of the string due to the cyclical symmetry of the partial amplitudes,
\ba
\bs{J}_{n}^{(1)}(1_{g},2_{g},3_{g},\cdots,(n-1)_{g},n_{g})&=&\bs{J}_{2}^{(1)}(1_{g},2_{g})+\bs{J}_{2}^{(1)}(2_{g},3_{g})+\cdots\nn\\
&+&\bs{J}_{2}^{(1)}((n-1)_{g},n_{g})+\bs{J}_{2}^{(1)}(n_{g},1_{g}).\label{eq:jsum}
\ea
The double unresolved integrated dipoles are given by,
\ba
\bs{J}_{2}^{(2)}(1_{g},2_{g})&=&\frac{1}{4}{\cal{F}}_{4}^{0}(s_{{1}{2}})+\frac{1}{3}{\cal{F}}_{3}^{1}(s_{{1}{2}})+\frac{1}{3}\frac{b_{0}}{\epsilon}{\cal{F}}_{3}^{0}(s_{{1}{2}})\Big[\biggl(\frac{|s_{{1}{2}}|}{\mu^{2}}\biggr)^{-\epsilon}-1\Big]\nn\\
&-&\frac{1}{9}\big[{\cal{F}}_{3}^{0}\otimes{\cal{F}}_{3}^{0}\big](s_{12}),\\
\bs{J}_{2}^{(2)}(\hb{1}_{g},2_{g})&=&\frac{1}{2}{\cal{F}}_{4,g}^{0}(s_{\b{1}{2}})+\frac{1}{2}{\cal{F}}_{3,g}^{1}(s_{\b{1}{2}})+\frac{1}{2}\frac{b_{0}}{\epsilon}{\cal{F}}_{3,g}^{0}(s_{\b{1}{2}})\Big[\biggl(\frac{|s_{\b{1}{2}}|}{\mu^{2}}\biggr)^{-\epsilon}-1\Big]\nn\\
&-&\frac{1}{4}\big[{\cal{F}}_{3,g}^{0}\otimes{\cal{F}}_{3,g}^{0}\big](s_{\b{1}2})-\frac{1}{2}\overline{\Gamma}_{gg}^{(2)}(x_{1})\delta(1-x_{2}),\\
\bs{J}_{2}^{(2)}(\hb{1}_{g},\hb{2}_{g})&=&{\cal{F}}_{4,gg}^{0,\rm{adj}}(s_{\b{1}\b{2}})+\frac{1}{2}{\cal{F}}_{4,gg}^{0,\rm{n.adj}}(s_{\b{1}\b{2}})+{\cal{F}}_{3,gg}^{1}(s_{\b{1}\b{2}})+\frac{b_{0}}{\epsilon}{\cal{F}}_{3,gg}^{0}(s_{\b{1}\b{2}})\Big[\biggl(\frac{|s_{\b{1}\b{2}}|}{\mu^{2}}\biggr)^{-\epsilon}-1\Big]\nn\\
&-&\big[{\cal{F}}_{3,gg}^{0}\otimes{\cal{F}}_{3,gg}^{0}\big](s_{\b{1}\b{2}})-\frac{1}{2}\overline{\Gamma}_{gg;gg}^{(2)}(x_{1},x_{2}),
\ea
where the relevant mass factorization kernels are defined by~\cite{GehrmannDeRidder:2012dg},
\ba
\overline{\Gamma}_{gg;gg}^{(2)}(x_{1},x_{2})&=&\overline{\Gamma}_{gg}^{(2)}(x_{1})\delta(1-x_{2})+\overline{\Gamma}_{gg}^{(2)}(x_{2})\delta(1-x_{1}),\\
\overline{\Gamma}_{gg}^{(2)}(x_{1})&=&-\frac{1}{2\e}\Big(p_{gg}^{1}(x_{1})+\frac{\beta_{0}}{\e}p_{gg}^{0}(x_{1})\Big).\label{eq:gamma2b}
\ea
Using the integrated dipoles, and evaluating the colour charge sandwiches directly, allows the pole structure of one- and two-loop contributions to the cross section to be written in terms of single and double unresolved integrated antenna dipoles. The initial-final and initial-initial dipoles contain mass factorization kernels, however as stated in Eqs.~\eqref{eq:slcmf1} and~\eqref{eq:slcmf2}, the mass factorization contribution is zero at sub-leading colour so all mass factorization kernels in the integrated dipoles ultimately cancel in the full subtraction term. The pole cancellation with the relevant subtraction terms can then be achieved in a clear and simple fashion as we will show in Secs.~\ref{sec:RV} and~\ref{sec:VV}.

\section{Double-real contribution}
\label{sec:RR}

The double real six gluon tree-level amplitude squared is given by,
\begin{eqnarray}
&&\bs{A}_{6}^{0}(\{p\})=g^{8}N^{4}(N^2-1)\bigg\{
\sum_{\sigma\in S_{6}/Z_{6}}A_{6}^{0}(1,\sigma(2),\sigma(3),\sigma(4),\sigma(5),\sigma(6))\nn\\
&&+\frac{2}{N^2}\ {\cal A}_{6}^{0\dagger}(1,\sigma(2),\sigma(3),\sigma(4),\sigma(5),\sigma(6))
\Big[{\cal A}_{6}^{0}(1,\sigma(3),\sigma(5),\sigma(2),\sigma(6),\sigma(4))\nn\\
&&+{\cal A}_{6}^{0}(1,\sigma(3),\sigma(6),\sigma(4),\sigma(2),\sigma(5))
+{\cal A}_{6}^{0}(1,\sigma(4),\sigma(2),\sigma(6),\sigma(3),\sigma(5))\Big]
\bigg\},
\end{eqnarray}
where the sum $S_{6}/Z_{6}$ is the group of 5! non-cyclic permutations of the six gluons.

The leading colour contribution and the double real subtraction term has been discussed in Ref.~\cite{Glover:2010im}.  At sub-leading colour, the double real radiation contribution is given by,
 \begin{eqnarray}
\lefteqn{{\rm d}\hat\sigma_{NNLO}^{RR}= {\cal N}_{LO} \left(\frac{\alpha_s}{2\pi}\right)^2\frac{\b{C}(\e)^2}{C(\e)^2}
{\rm d}\Phi_{4}(p_{3},\ldots,p_{6};p_1;p_2)\, \JET_{2}^{(4)}(p_{3},\ldots,p_{6})
 \, \frac{2}{4!} \sum_{\sigma\in S_{6}/Z_{6}}}\nn\\
&\times &{\cal A}_{6}^{0\dagger}(1,\sigma(2),\sigma(3),\sigma(4),\sigma(5),\sigma(6))
\Big[{\cal A}_{6}^{0}(1,\sigma(3),\sigma(5),\sigma(2),\sigma(6),\sigma(4))\nn\\
&&+\ {\cal A}_{6}^{0}(1,\sigma(3),\sigma(6),\sigma(4),\sigma(2),\sigma(5)
+{\cal A}_{6}^{0}(1,\sigma(4),\sigma(2),\sigma(6),\sigma(3),\sigma(5))\Big],
\label{eq:RR}
\end{eqnarray}
where $\b{C}(\e)=8\pi^2 C(\e)=(4\pi)^\e e^{-\e\gamma}$ and the overall factor is given by,
\ba
{\cal{N}}_{LO}&=&\frac{1}{2s}\frac{1}{4(N^2-1)^2}(g^2 N)^2(N^2-1).
\ea
In Eq.~\eqref{eq:RR} we can see that the tree-level six gluon squared matrix element at sub-leading colour can be written as three incoherent interferences, summed over permutations.  The three orderings,
\ba
&&{\cal{A}}_{6}^{0}(1,\sigma{(3)},\sigma{(5)},\sigma{(2)},\sigma{(6)},\sigma{(4)}),\nn\\
&&{\cal{A}}_{6}^{0}(1,\sigma{(3)},\sigma{(6)},\sigma{(4)},\sigma{(2)},\sigma{(5)}),\nn\\
&&{\cal{A}}_{6}^{0}(1,\sigma{(4)},\sigma{(2)},\sigma{(6)},\sigma{(3)},\sigma{(5)}),\nn
\ea
are the only independent orderings that exist for six gluon scattering which have no common neighbouring pairs of partons with the conjugate amplitude's ordering,
\ba
&&{\cal{A}}_{6}^{0,\dagger}(1,\sigma{(2)},\sigma{(3)},\sigma{(4)},\sigma{(5)},\sigma{(6)}).\nn
\ea
One immediate consequence of this is that the sub-leading colour matrix element does not contain any single, double or triple collinear collinear divergences.  With no collinear divergences present in the double real cross section, the only divergences to be removed are those associated with single and double soft gluons.

The double real subtraction term can be divided into five distinct contributions,
\ba
\dsigma_{NNLO}^{S}&=&\dsigma_{NNLO}^{S,a}+\dsigma_{NNLO}^{S,b}+\dsigma_{NNLO}^{S,c}+\dsigma_{NNLO}^{S,d}+\dsigma_{NNLO}^{S,e},
\ea
which will be discussed in the following sections.

\subsection{Single unresolved subtraction term}
\label{sec:ssRR}

The interferences in Eq.\eqref{eq:RR} contain no collinear divergences but do contain soft singularities.  In the single soft limit the colour ordered partial amplitudes factorize~\cite{Berends:1988zn,Catani:1999ss},
\begin{equation}
{\cal M}_{n+1}^0(\hdots,p_i,p_j,p_k,\hdots)\stackrel{j\to0}{\longrightarrow}\S^{0}(p_i,p_j,p_k){\cal M}_{n}^0(\hdots,p_i,p_k,\hdots),
\end{equation}
where 
\ba
\S^{0}(p_i,p_j,p_k)&=&S_{\mu}^{0}(p_i,p_j,p_k)\epsilon^{\mu}(p_j).
\ea
The tree-level single soft current is given by\cite{Weinzierl:2003fx},
\begin{equation}
S_{\mu}^{0}(p_i,p_j,p_k)=2\frac{p_{i}^{\rho}F_{\rho\mu\sigma}(p_j)p_{k}^{\sigma}}{s_{ij}s_{jk}},
\end{equation}
and $F_{\rho\mu\sigma}(p)$ is defined by,
\begin{equation}
F_{\rho\mu\sigma}(p)=g_{\rho\mu}p_{\sigma}-p_{\rho}g_{\mu\sigma}\;.
\end{equation}
Summing over polarizations allows the soft limit of the  interference to be written in terms of eikonal factors,
\ba
\lefteqn{{\cal{A}}_{6}^{0,\dagger}(\cdots,a,i,b,\cdots){\cal{A}}_{6}^{0}(\cdots,c,i,d,\cdots)\stackrel{i\to0}{\longrightarrow}}\nn\\
&&\frac{1}{2}\Big[S_{aid}+S_{bic}-S_{aic}-S_{bid}\Big]{\cal{A}}_{5}^{0,\dagger}(\cdots,a,{b},\cdots){\cal{A}}_{5}^{0}(\cdots,c,{d},\cdots),\label{eq:fact}
\ea
where the eikonal factor is,
\ba
S_{ijk}&=&\frac{2s_{ik}}{s_{ij}s_{jk}}.
\ea
The eikonal factors have uniquely defined hard radiators and can be immediately promoted to antenna functions with an appropriate momentum mapping~\cite{GehrmannDeRidder:2005cm,Daleo:2006xa} to obtain a candidate subtraction term for the single soft limit of a generic tree-level interference,
\ba
\lefteqn{{\cal{A}}_{6}^{0,\dagger}(\cdots,a,i,b,\cdots){\cal{A}}_{6}^{0}(\cdots,c,i,d,\cdots)\stackrel{i\to0}{\approx}}\nn\\
\frac{1}{2}&\Big[&X_{3}^{0}(a,i,d)\ {\cal{A}}_{5}^{0,\dagger}(\cdots,\wt{(ai)},{b},\cdots){\cal{A}}_{5}^{0}(\cdots,{c},\wt{(id)},\cdots)\nn\\
&+&X_{3}^{0}(b,i,c)\ {\cal{A}}_{5}^{0,\dagger}(\cdots,{a},\wt{(bi)},\cdots){\cal{A}}_{5}^{0}(\cdots,\wt{(ic)},{d},\cdots)\nn\\
&-&X_{3}^{0}(a,i,c)\ {\cal{A}}_{5}^{0,\dagger}(\cdots,\wt{(ai)},{b},\cdots){\cal{A}}_{5}^{0}(\cdots,\wt{(ic)},{d},\cdots)\nn\\
&-&X_{3}^{0}(b,i,d)\ {\cal{A}}_{5}^{0,\dagger}(\cdots,{a},\wt{(bi)},\cdots){\cal{A}}_{5}^{0}(\cdots,{c},\wt{(id)},\cdots)\Big].\label{eq:fact2}
\ea
For convenience the momentum mapping shown in Eq.~\eqref{eq:fact2} is a final-final type but the factorization pattern is true for all mappings. It can be easily seen that the collinear limits of the antennae in this block of terms cancel in Eq.~\eqref{eq:fact2}.

The single unresolved subtraction term is given by,
\begin{eqnarray}
{\rm d}\hat\sigma_{NNLO}^{S,a}&=&{\cal N}_{LO} \left(\frac{\alpha_s }{2\pi}\right)^2
{\rm d}\Phi_{4}(p_{3},\ldots,p_{6};p_1;p_2)\,
 \, \frac{12}{4!} \sum_{(i,j,k,l)\in P(3,4,5,6)}
2{\rm{Re}}\Big\{\nonumber\\
&-& F_{3}^{0}(\h{1},l,\h{2}) {\cal{A}}_{5}^{0}(\hb{1},\hb{2},\t{i},\t{j},\t{k}){\cal{A}}_{5}^{0,\dagger}(\hb{1},\t{j},\hb{2},\t{k},\t{i})J_{2}^{(3)}(p_{\t{i}},p_{\t{j}},p_{\t{k}})\nn\\
&-& f_{3}^{0}(\h{2},l,i){\cal{A}}_{5}^{0}(\h{1},\hb{2},\wt{(li)},j,k){\cal{A}}_{5}^{0,\dagger}(\h{1},j,\hb{2},k,\wt{(li)})J_{2}^{(3)}(p_j,p_k,p_{\wt{(li)}})\nn\\
&-&f_{3}^{0}(i,l,j){\cal{A}}_{5}^{0}(\h{1},\h{2},\wt{(il)},\wt{(lj)},k){\cal{A}}_{5}^{0,\dagger}(\h{1},\wt{(lj)},\h{2},k,\wt{(il)})J_{2}^{(3)}(p_k,p_{\wt{(il)}},p_{\wt{(lj)}})\nn\\
&-&f_{3}^{0}(j,l,k){\cal{A}}_{5}^{0}(\h{1},\h{2},i,\wt{(jl)},\wt{(lk)}){\cal{A}}_{5}^{0,\dagger}(\h{1},\wt{(jl)},\h{2},\wt{(lk)},i)J_{2}^{(3)}(p_i,p_{\wt{(jl)}},p_{\wt{(lk)}})\nn\\
&-& f_{3}^{0}(\h{1},l,k){\cal{A}}_{5}^{0}(\hb{1},\h{2},i,j,\wt{(lk)}){\cal{A}}_{5}^{0,\dagger}(\hb{1},j,\h{2},\wt{(lk)},i)J_{2}^{(3)}(p_i,p_j,p_{\wt{(lk)}})\nn\\
&+& f_{3}^{0}(\h{1},l,j){\cal{A}}_{5}^{0}(\hb{1},\h{2},i,\wt{(lj)},k){\cal{A}}_{5}^{0,\dagger}(\hb{1},\wt{(lj)},\h{2},k,i)J_{2}^{(3)}(p_i,p_k,p_{\wt{(lj)}})\nn\\
&+& f_{3}^{0}(\h{2},l,j){\cal{A}}_{5}^{0}(\h{1},\hb{2},i,\wt{(lj)},k){\cal{A}}_{5}^{0,\dagger}(\h{1},\wt{(lj)},\hb{2},k,i)J_{2}^{(3)}(p_i,p_k,p_{\wt{(lj)}})\nn\\
&+& f_{3}^{0}(\h{2},l,k){\cal{A}}_{5}^{0}(\h{1},\hb{2},i,j,\wt{(lk)}){\cal{A}}_{5}^{0,\dagger}(\h{1},j,\hb{2},\wt{(lk)},i)J_{2}^{(3)}(p_i,p_j,p_{\wt{(lk)}})\nn\\
&+&f_{3}^{0}(i,l,k){\cal{A}}_{5}^{0}(\h{1},\h{2},\wt{(il)},j,\wt{(lk)}){\cal{A}}_{5}^{0,\dagger}(\h{1},j,\h{2},\wt{(lk)},\wt{(il)})J_{2}^{(3)}(p_j,p_{\wt{(il)}},p_{\wt{(lk)}})\nn\\
&+& f_{3}^{0}(\h{1},l,i){\cal{A}}_{5}^{0}(\hb{1},\h{2},\wt{(li)},j,k){\cal{A}}_{5}^{0,\dagger}(\hb{1},j,\h{2},k,\wt{(li)})J_{2}^{(3)}(p_j,p_k,p_{\wt{(li)}})\Big\}.
\label{eq:siga2}
\end{eqnarray}
Once analytically integrated, Eq.~\eqref{eq:siga2} is added back as part of the real-virtual subtraction term where it cancels the explicit IR poles of the real-virtual matrix element.

\subsection{Double unresolved subtraction term}

The only double unresolved divergences present are those associated with two simultaneously soft gluons.  In the double soft gluon limit, the full squared gluonic matrix element factorizes in the following way~\cite{Catani:1999ss},
\ba
\bs{A}_{6}^{0}(\{p\})&\stackrel{i,j\to0}{\propto}&\sum_{(a,b)}\sum_{(c,d)}S_{aib}S_{cjd}~\la{\cal{A}}_{4}^{0}|(\bt_{a}\cdot\bt_{b})(\bt_{c}\cdot\bt_{d})|{\cal{A}}_{4}^{0}\ra\nn\\
&-&N~\sum_{(a,b)}S_{ab}(i,j)~\la{\cal{A}}_{4}^{0}|\bt_{a}\cdot\bt_{b}|{\cal{A}}_{4}^{0}\ra,\label{eq:2gsoftfac}
\ea
where the four parton double soft function, $S_{ab}(i,j)$~\cite{Catani:1999ss} is related to the double soft function, $S_{aijb}$, derived in~\cite{Campbell:1997hg}.  The last term in Eq.~\eqref{eq:2gsoftfac} is proportional to the sandwich $\la{\cal{A}}_{4}^{0}|\bt_{a}\cdot\bt_{b}|{\cal{A}}_{4}^{0}\ra$, which as stated in Eq.~\eqref{eq:slczero}, does not contribute to the sub-leading colour contribution. Accordingly, we find that the double soft factorization pattern involves only eikonal factors, e.g. in the limit where gluons five and six go simultaneously soft,
\ba
\lefteqn{{\cal{SLC}}\Big(\bs{A}_{6}^{0}(\{p\})\Big)\stackrel{5,6\to0}{\longrightarrow}}\nn\\
\frac{1}{4}&\Big[&\Big(S_{153}+S_{254}-S_{154}-S_{254}\Big)\Big(S_{163}+S_{264}-S_{162}-S_{364}\Big) A_{4}^{0}(\hat{1},\hat{2},3,4)\nn\\
&+&\Big(S_{153}+S_{254}-S_{154}-S_{254}\Big)\Big(S_{162}+S_{364}-S_{164}-S_{263}\Big) A_{4}^{0}(\hat{1},\hat{2},4,3)\nn\\
&+&\Big(S_{152}+S_{354}-S_{154}-S_{253}\Big)\Big(S_{162}+S_{364}-S_{163}-S_{264}\Big)A_{4}^{0}(\hat{1},3,\hat{2},4)\Big].\label{eq:2gsoftfac2}\nn\\
\ea
We can also study the double soft limit of the partial amplitudes directly~\cite{Berends:1988zn,Catani:1999ss},
\begin{equation}
{\cal M}_{n+2}^0(\hdots,p_i,p_j,p_k,p_l,\hdots)\stackrel{j,k\to0}{\longrightarrow}\S^{0}(p_i,p_j,p_k,p_l){\cal M}_{n}^0(\hdots,p_i,p_l,\hdots),
\end{equation}
where
\ba
\S^{0}(p_i,p_j,p_k,p_l)&=&S_{\mu\nu}(p_i,p_j,p_k,p_l)\epsilon^{\mu}(p_j)\epsilon^{\nu}(p_k),
\ea
and the double soft current can be written as~\cite{Weinzierl:2003fx},
\begin{eqnarray}
\lefteqn{S_{\mu\nu}(p_i,p_j,p_k,p_l)=}\nn\\
&&4\Big[\frac{p_{i}^{\rho}F_{\rho\mu}^{\sigma}(p_j)F_{\sigma\nu\tau}(p_k)p_{l}^{\tau}}{s_{ij}s_{jk}s_{kl}}
-\frac{p_{i}^{\rho}F_{\rho\mu}^{\sigma}(p_j)F_{\sigma\nu\tau}(p_k)p_{i}^{\tau}}{s_{ij}s_{jk}(s_{ij}+s_{ik})}
-\frac{p_{l}^{\rho}F_{\rho\mu}^{\sigma}(p_j)F_{\sigma\nu\tau}(p_k)p_{l}^{\tau}}{s_{jk}s_{kl}(s_{jl}+s_{kl})}\Big]\;.
\label{eq:dsoftcurr}
\end{eqnarray}
When taking the double soft limit of Eq.~\eqref{eq:RR}, we encounter contractions between single and double soft currents. Summing over all colour orderings, we obtain symmetric sums of double soft currents which can be rewritten using the identity~\cite{Berends:1988zn},
\begin{equation}
S_{\mu\nu}(p_i,p_j,p_k,p_l)+S_{\nu\mu}(p_i,p_k,p_j,p_l)=S_{\mu}(p_i,p_j,p_l)S_{\nu}(p_i,p_k,p_l)\;.
\label{eq:softcur1}
\end{equation}
Any remaining terms involving the double soft current can be eliminated using,
\begin{eqnarray}
-S_{\mu\nu}(p_i,p_j,p_k,p_l)&+&S_{\mu\nu}(p_i,p_j,p_k,p_a)+S_{\nu\mu}(p_l,p_k,p_j,p_c)-S_{\nu\mu}(p_a,p_k,p_j,p_c)\nn\\
&=&-S_{\mu}(p_i,p_j,p_k)S_{\nu}(p_i,p_k,p_l)+S_{\mu}(p_i,p_j,p_k)S_{\nu}(p_i,p_k,p_a)\nn\\
&&-S_{\mu}(p_k,p_j,p_c)S_{\nu}(p_c,p_k,p_l)+S_{\mu}(p_k,p_j,p_c)S_{\nu}(p_c,p_k,p_a),\label{eq:softcur2}
\end{eqnarray}
so that the double soft limit of the sub-leading colour matrix element can be written purely in terms of eikonal factors, as described in Eq.~\eqref{eq:2gsoftfac2}. To the best of our knowledge, the relation in Eq.~\eqref{eq:softcur2} does not exist in the literature and can be confirmed analytically using~\eqref{eq:dsoftcurr}.   The resulting subtraction terms can be obtained by promoting each eikonal factor to a three-parton tree-level antenna, as outlined in Sec.~\ref{sec:ssRR}.

At leading colour, the double unresolved subtraction term is partitioned into three terms depending on the \emph{colour connection} of the unresolved partons:
\begin{itemize}
\item Colour connected, where the unresolved partons, $i,j$, are colour connected to each other and a single pair of neighbouring hard radiators, $a,b$, corresponding to the colour ordering $(\cdots,a,i,j,b,\cdots)$.
\item Almost colour connected, where the unresolved partons, $i,j$, are not colour connected but are each colour connected to a neighbouring pair of hard radiators, $a,c$ and $c,b$, with one hard radiator in common, corresponding to the colour ordering $(\cdots,a,i,c,j,b,\cdots)$.
\item Colour disconnected, where the unresolved partons, $i,j$, are not colour connected and have no hard radiating neighbours in common, corresponding to the colour ordering $(\cdots,a,i,b,\cdots,c,j,d,\cdots)$.
\end{itemize}
This classification is particularly useful for leading colour calculations because, due to colour coherence, the leading colour cross section is formed from squared partial amplitudes, each with a definite colour ordering.  

For sub-leading colour calculations this distinction is not so apparent. An incoherent interference generally produces a complicated factorization pattern, particularly in the soft limits, and does not have a simple connection with the colour ordering inherent to squared partial amplitudes. For such interferences, what is colour connected, almost colour connected or colour disconnected is not immediately obvious. We have already established that the double unresolved subtraction term consists of iterated three-parton antennae.  For the remainder of this paper we consider the slightly looser definitions of terms:
\begin{itemize}
\item $\dsigma_{NNLO}^{S,b}$, the antennae have repeated hard radiators, $\sim X_{3}^{0}(a,i,b)X_{3}^{0}(A,j,B)$,
\item $\dsigma_{NNLO}^{S,c}$, the antennae have one repeated radiator, $\sim X_{3}^{0}(a,i,c)X_{3}^{0}(C,j,b)$,
\item $\dsigma_{NNLO}^{S,d}$, the antennae have no repeated radiators, $\sim X_{3}^{0}(a,i,b)X_{3}^{0}(c,j,d)$,
\end{itemize}
where $A, B, C$ denote the repeated hard radiators which have composite momenta resulting from the appropriate momentum mapping fixed by the primary antenna, e.g. $(a,i,b)\to(A,B)$.  The subtraction terms, $\dsigma_{NNLO}^{S,b}$ and $\dsigma_{NNLO}^{S,c}$ are integrated and added back to the real-virtual cross section while  $\dsigma_{NNLO}^{S,d}$ can be simultaneously integrated over both unresolved partons and so is added back in the double virtual subtraction term.

The sum of the three double unresolved subtraction terms is given by,
\begin{eqnarray}
\lefteqn{\dsigma_{NNLO}^{S,b}+\dsigma_{NNLO}^{S,c}+\dsigma_{NNLO}^{S,d}={\cal N}_{LO} \left(\frac{\alpha_s}{2\pi}\right)^2\frac{\b{C}(\e)^2}{C(\e)^2}
{\rm d}\Phi_{4}(p_{3},\ldots,p_{6};p_1;p_2)\frac{12}{4!}\sum_{(i,j,k,l)\in P(3,4,5,6)}}\nn\\
\Big\{&-&\frac{1}{2}F_3^0(\h{1},l,\h{2})F_3^0(\h{\b{1}},\t{k},\h{\b{2}}) A_4^0(\h{\bb{1}},\t{\t{j}},\h{\bb{2}},\t{\t{i}})J_{2}^{(2)}(p_{\t{\t{j}}},p_{\t{\t{i}}})\nn\\
&-&f_3^0(\h{1},l,i)f_3^0(\h{\b{1}},k,\wt{(li)}) A_4^0(\h{\bb{1}},\h{2},\wt{(k\wt{(li)})},j)J_{2}^{(2)}(p_{\wt{(k\wt{(li)})}},p_{j})\nn\\
&-&f_3^0(2,l,i)f_3^0(\h{\b{2}},k,\wt{(li)}) A_4^0(\h{1},\h{\bb{2}},j,\wt{(k\wt{(li)})})J_{2}^{(2)}(p_{j},p_{\wt{(k\wt{(li)})}})\nn\\
&-&\frac{1}{2}f_3^0(i,l,j)f_3^0(\wt{(il)},k,\wt{(lj)})A_4^0(\h{1},\wt{(\wt{(il)}k)},\h{2},\wt{(k\wt{(lj)})}) J_{2}^{(2)}(p_{\wt{(\wt{(il)}k)}},p_{\wt{(k\wt{(lj)})}})\nn\\\nn\\
&+&\frac{1}{2}F_3^0(\h{1},l,\h{2})f_3^0(\h{\b{1}},\t{k},\t{j})\nn\\
&&\times\Big[A_4^0(\h{\bb{1}},\h{\b{2}},\wt{(\t{k}\t{j})},\t{i})+A_4^0(\h{\bb{1}},\wt{(\t{k}\t{j})},\h{\b{2}},\t{i})- A_4^0(\h{\bb{1}},\h{\b{2}},\t{i},\wt{(\t{k}\t{j})})\Big]J_{2}^{(2)}(p_{\t{i}},p_{\wt{(\t{k}\t{j})}})\nn\\
&+&\frac{1}{2}F_3^0(\h{1},l,\h{2})f_3^0(\h{\b{2}},\t{k},\t{j})\nn\\
&&\times\Big[A_4^0(\h{\b{1}},\h{\bb{2}},\t{i},\wt{(\t{k}\t{j})})+A_4^0(\h{\b{1}},\wt{(\t{k}\t{j})},\h{\bb{2}},\t{i})-A_4^0(\h{\b{1}},\h{\bb{2}},\wt{(\t{k}\t{j})},\t{i})\Big]J_{2}^{(2)}(p_{\wt{(\t{k}\t{j})}},p_{\t{i}})\nn\\
&+&\frac{1}{2}f_3^0(\h{1},l,i)f_3^0(j,k,\wt{(li)})\nn\\
&&\times\Big[A_4^0(\h{\b{1}},\h{2},\wt{(k\wt{(li)})},\wt{(jk)})+A_4^0(\h{\b{1}},\wt{(jk)},\h{2},\wt{(k\wt{(li)})})-A_4^0(\h{\b{1}},\h{2},\wt{(jk)},\wt{(k\wt{(li)})})\Big]J_{2}^{(2)}(p_{\wt{(jk)}},p_{\wt{(k\wt{(li)})}})\nn\\
&+&\frac{1}{2}f_3^0(\h{1},l,i)f_3^0(\h{2},k,\wt{(li)})\nn\\
&&\times\Big[A_4^0(\h{\b{1}},\h{\b{2}},j,\wt{(k\wt{(li)})})+A_4^0(\h{\b{1}},\h{\b{2}},\wt{(k\wt{(li)})},j)-A_4^0(\h{\b{1}},\wt{(k\wt{(li)})},\h{\b{2}},j)\Big]J_{2}^{(2)}(p_{\wt{(k\wt{(li)})}},p_{j})\nn\\
&+&\frac{1}{2}f_3^0(\h{1},l,i)f_3^0(\h{\b{1}},k,j)\nn\\
&&\times\Big[A_4^0(\h{\bb{1}},\h{2},\wt{(kj)},\wt{(li)})+A_4^0(\h{\bb{1}},\h{2},\wt{(li)},\wt{(kj)})-A_4^0(\h{\bb{1}},\wt{(kj)},\h{2},\wt{(li)})\Big]J_{2}^{(2)}(p_{\wt{(kj)}},p_{\wt{(li)}})\nn\\
&+&\frac{1}{2}f_3^0(\h{1},l,i)F_3^0(\h{\b{1}},k,\h{2})\nn\\
&&\times\Big[A_4^0(\h{\bb{1}},\h{\b{2}},\wt{(\wt{(li)})},\t{j})+A_4^0(\h{\bb{1}},\wt{(\wt{(li)})},\h{\b{2}},\t{j})-A_4^0(\h{\bb{1}},\h{\b{2}},\tilde{j},\wt{(\wt{(li)})})\Big]J_{2}^{(2)}(p_{\t{j}},p_{\wt{(\wt{(li)})}})\nn\\
&+&\frac{1}{2}f_3^0(2,l,i)f_3^0(j,k,\wt{(li)})\nn\\
&&\times\Big[A_4^0(\h{1},\h{\b{2}},\wt{(jk)},\wt{(k\wt{(li)})})+A_4^0(\h{1},\wt{(jk)},\h{\b{2}},\wt{(k\wt{(li)})})-A_4^0(\h{1},\h{\b{2}},\wt{(k\wt{(li)})},\wt{(jk)})\Big]J_{2}^{(2)}(p_{\wt{(k\wt{(li)})}},p_{\wt{(jk)}})\nn\\
&+&\frac{1}{2}f_3^0(2,l,i)f_3^0(\h{1},k,\wt{(li)}) \nn\\
&&\times\Big[A_4^0(\h{\b{1}},\h{\b{2}},j,\wt{(k\wt{(li)})})+A_4^0(\h{\b{1}},\h{\b{2}},\wt{(k\wt{(li)})},j)-A_4^0(\h{\b{1}},\wt{(k\wt{(li)})},\h{\b{2}},j)\Big]J_{2}^{(2)}(p_{\wt{(k\wt{(li)})}},p_{j})\nn\\
&+&\frac{1}{2}f_3^0(2,l,i)f_3^0(\h{\b{2}},k,j) \nn\\
&&\times\Big[A_4^0(\h{1},\h{\bb{2}},\wt{(kj)},\wt{(li)})A_4^0(\h{1},\h{\bb{2}},\wt{(li)},\wt{(kj)})-A_4^0(\h{1},\wt{(kj)},\h{\bb{2}},\wt{(li)})\Big]J_{2}^{(2)}(p_{\wt{(kj)}},p_{\wt{(li)}})\nn\\
&+&\frac{1}{2}f_3^0(2,l,i)F_3^0(\h{1},k,\h{\b{2}}) \nn\\
&&\times\Big[A_4^0(\h{\b{1}},\h{\bb{2}},\t{j},\wt{(\wt{(li)})})+A_4^0(\h{\b{1}},\wt{(\wt{(li)})},\h{\bb{2}},\t{j})-A_4^0(\h{\b{1}},\h{\bb{2}},\wt{(\wt{(li)})},\t{j})\Big]J_{2}^{(2)}(p_{\t{j}},p_{\wt{(\wt{(li)})}})\nn\\
&+&\frac{1}{4}f_3^0(i,l,j)f_3^0(\h{1},k,\wt{(il)})\nn\\
&&\times\Big[A_4^0(\h{\b{1}},\h{2},\wt{(k\wt{(il)})},\wt{(lj)})+A_4^0(\h{\b{1}},\wt{(k\wt{(il)})},\h{2},\wt{(lj)})-A_4^0(\h{\b{1}},\h{2},\wt{(lj)},\wt{(k\wt{(il)})})\Big]J_{2}^{(2)}(p_{\wt{(k\wt{(il)})}},p_{\wt{(lj)}})\nn\\
&+&\frac{1}{4}f_3^0(i,l,j)f_3^0(\h{1},k,\wt{(lj)})\nn\\
&&\times\Big[A_4^0(\h{\b{1}},\h{2},\wt{(k\wt{(lj)})},\wt{(il)})+A_4^0(\h{\b{1}},\wt{(il)},\h{2},\wt{(k\wt{(lj)})})-A_4^0(\h{\b{1}},\h{2},\wt{(il)},\wt{(k\wt{(lj)})})\Big]J_{2}^{(2)}(p_{\wt{(il)}},p_{\wt{(k\wt{(lj)})}})\nn\\
&+&\frac{1}{4}f_3^0(i,l,j)f_3^0(\h{2},k,\wt{(il)})\nn\\
&&\times\Big[A_4^0(\h{1},\h{\b{2}},\wt{(lj)},\wt{(k\wt{(il)})})+A_4^0(\h{1},\wt{(k\wt{(il)})},\h{\b{2}},\wt{(lj)})-A_4^0(\h{1},\h{\b{2}},\wt{(k\wt{(il)})},\wt{(lj)})\Big]J_{2}^{(2)}(p_{\wt{(k\wt{(il)})}},p_{\wt{(lj)}})\nn\\
&+&\frac{1}{4}f_3^0(i,l,j)f_3^0(\h{2},k,\wt{(lj)})\nn\\
&&\times\Big[A_4^0(\h{1},\h{\b{2}},\wt{(il)},\wt{(k\wt{(lj)})})+A_4^0(\h{1},\wt{(il)},\h{\b{2}},\wt{(k\wt{(lj)})})-A_4^0(\h{1},\h{\b{2}},\wt{(k\wt{(lj)})},\wt{(il)})\Big]J_{2}^{(2)}(p_{\wt{(k\wt{(lj)})}},p_{\wt{(il)}})\nn\\
\nn\\
&-&F_3^0(\h{1},l,\h{2})f_3^0(\tilde{j},\tilde{k},\t{i}) A_4^0(\h{\b{1}},\wt{(\tilde{j}\tilde{k})},\h{\b{2}},\wt{(\tilde{k}\tilde{i})})J_{2}^{(2)}(p_{\wt{(\tilde{j}\tilde{k})}},p_{\wt{(\tilde{k}\tilde{i})}})\nn\\
&-&f_3^0(\h{1},l,i)f_3^0(\h{2},k,j) A_4^0(\h{\b{1}},\h{\b{2}},\wt{(li)},\wt{(kj)})J_{2}^{(2)}(p_{\wt{(li)}},p_{\wt{(kj)}})\Big\}.
\end{eqnarray}

In single soft limits, the single and double unresolved subtraction terms $\dsigma_{NNLO}^{S,a,b,c,d}$ over-subtract the divergences of the matrix element, and so a large angle soft subtraction term is introduced to compensate for this over-subtraction, denoted by $\dsigma_{NNLO}^{S,e}$,
\begin{eqnarray}
&&\dsigma_{NNLO}^{S,e}={\cal N}_{LO} \left(\frac{\alpha_s}{2\pi}\right)^2\frac{\b{C}(\e)^2}{C(\e)^2}
{\rm d}\Phi_{4}(p_{3},\ldots,p_{6};p_1;p_2)\frac{12}{4!} \sum_{(i,j,k,l)\in P(3,4,5,6)}\Big\{\nn\\
&&\frac{1}{4}  \Big(-S_{1l\wt{(il)}}+S_{1\t{l}\wt{(\wt{(il)})}}-S_{2l\wt{(il)}}+S_{2\t{l}\wt{(\wt{(il)})}}-S_{1l\wt{(lj)}}+S_{1\t{l}\wt{(\wt{(lj)})}}-S_{2l\wt{(lj)}}+S_{2\t{l}\wt{(\wt{(lj)})}}\nn\\
&&~-2 S_{1\t{l}2}+2 S_{1l2}\Big)~F_3^0(\h{1},k,\h{2}) A_4^0(\h{\b{1}},\wt{(\wt{(il)})},\h{\b{2}},\wt{(\wt{(lj)})})J_{2}^{(2)}(p_{\wt{(\wt{(il)})}},p_{\wt{(\wt{(lj)})}})\nn\\
&-&\frac{1}{2}\Big(S_{1l\wt{(il)}}-S_{1\t{l}\wt{(\wt{(il)})}}-S_{2l\wt{(il)}}+S_{2\t{l}\wt{(\wt{(il)})}}-S_{1l\wt{(lj)}}+S_{1\t{l}\wt{(\wt{(lj)})}}+S_{2l\wt{(lj)}}-S_{2\t{l}\wt{(\wt{(lj)})}}\Big)\nn\\
&&\times~F_3^0(\h{1},k,\h{2}) A_4^0(\h{\b{1}},\h{\b{2}},\wt{(\wt{(il)})},\wt{(\wt{(lj)})})J_{2}^{(2)}(p_{\wt{(\wt{(il)})}},p_{\wt{(\wt{(lj)})}})\nn\\
&-&\frac{1}{2} \Big(S_{\wt{(k\wt{(il)})}l\wt{(jl)}}-S_{\wt{(il)}l\wt{(lj)}}-S_{2l\wt{(k\wt{(il)})}}+S_{2l\wt{(il)}}\Big)\nn\\
&&\times~f_3^0(\h{1},k,\wt{(il)}) A_4^0(\h{\b{1}},2,\wt{(lj)},\wt{(k\wt{(il)})})J_{2}^{(2)}(p_{\wt{(lj)}},p_{\wt{(k\wt{(il)})}})\nn\\
&+&\frac{1}{2}\Big(S_{\wt{(k\wt{(il)})}l\wt{(lj)}}-S_{\wt{(il)}l\wt{(lj)}}-S_{2l\wt{(k\wt{(il)})}}+S_{2l\wt{(il)}}\Big)\nn\\
&&\times~f_3^0(\h{1},k,\wt{(il)}) A_4^0(\h{\b{1}},\wt{(k\wt{(il)})},2,\wt{(lj)})J_{2}^{(2)}(p_{\wt{(lj)}},p_{\wt{(k\wt{(il)})}})\nn\\
&+&\frac{1}{2}\Big(S_{\wt{(k\wt{(il)})}l\wt{(lj)}}-S_{\wt{(il)}l\wt{(lj)}}-2 S_{1l\wt{(k\wt{(il)})}}+S_{2l\wt{(k\wt{(il)})}}+2 S_{1l\wt{(il)}}-S_{2l\wt{(il)}}\Big)\nn\\
&&\times~f_3^0(\h{1},k,\wt{(il)}) A_4^0(\h{\b{1}},2,\wt{(k\wt{(il)})},\wt{(lj)})J_{2}^{(2)}(p_{\wt{(lj)}},p_{\wt{(k\wt{(il)})}})\nn\\
&+& \frac{1}{2}\Big (S_{\wt{(k\wt{(jl)})}l\wt{(li)}}-S_{\wt{(jl)}l\wt{(li)}}-S_{1l\wt{(k\wt{(jl)})}}+S_{1l\wt{(jl)}}\Big)\nn\\
&&\times~f_3^0(\h{2},k,\wt{(jl)}) A_4^0(\h{1},\wt{(k\wt{(jl)})},\h{\b{2}},\wt{(li)})J_{2}^{(2)}(p_{\wt{(li)}},p_{\wt{(k\wt{(jl)})}})\nn\\
&-&\frac{1}{2}\Big(-S_{\wt{(k\wt{(jl)})}l\wt{(li)}}+S_{\wt{(jl)}l\wt{(li)}}-S_{1l\wt{(k\wt{(jl)})}}+2 S_{2l\wt{(k\wt{(jl)})}}+S_{1l\wt{(jl)}}-2 S_{2l\wt{(jl)}}\Big)\nn\\
&&\times~f_3^0(\h{2},k,\wt{(jl)}) A_4^0(\h{1},\h{\b{2}},\wt{(li)},\wt{(k\wt{(jl)})}) J_{2}^{(2)}(p_{\wt{(li)}},p_{\wt{(k\wt{(jl)})}})\nn\\
&-&\frac{1}{2}\Big(S_{\wt{(k\wt{(jl)})}l\wt{(li)}}-S_{\wt{(jl)}l\wt{(li)}}-S_{1l\wt{(k\wt{(jl)})}}+S_{1l\wt{(jl)}}\Big)\nn\\
&&\times~f_3^0(\h{2},k,\wt{(jl)}) A_4^0(\h{1},\h{\b{2}},\wt{(k\wt{(jl)})},\wt{(li)})J_{2}^{(2)}(p_{\wt{(li)}},p_{\wt{(k\wt{(jl)})}})\nn\\
&+&\frac{1}{4}\Big(S_{1l\wt{(\wt{(il)}k)}}-S_{2l\wt{(\wt{(il)}k)}}-S_{1l\wt{(il)}}+S_{2l\wt{(il)}}-S_{1l\wt{(k\wt{(lj)})}}+S_{2l\wt{(k\wt{(lj)})}}+S_{1l\wt{(lj)}}-S_{2l\wt{(lj)}}\Big)\nn\\
&&\times~f_3^0(\wt{(il)},k,\wt{(lj)}) A_4^0(\h{1},\h{2},\wt{(\wt{(il)}k)},\wt{(k\wt{(lj)})})J_{2}^{(2)}(p_{\wt{(\wt{(il)}k)}},p_{\wt{(k\wt{(lj)})}})\nn\\
&-&\frac{1}{4} \Big(S_{1l\wt{(\wt{(il)}k)}}-S_{2l\wt{(\wt{(il)}k)}}-S_{1l\wt{(il)}}+S_{2l\wt{(il)}}-S_{1l\wt{(k\wt{(lj)})}}+S_{2l\wt{(k\wt{(lj)})}}+S_{1l\wt{(lj)}}-S_{2l\wt{(lj)}}\Big)\nn\\
&&\times~f_3^0(\wt{(il)},k,\wt{(lj)})  A_4^0(\h{1},\h{2},\wt{(k\wt{(lj)})},\wt{(\wt{(il)}k)})J_{2}^{(2)}(p_{\wt{(\wt{(il)}k)}},p_{\wt{(k\wt{(lj)})}})\nn\\
&+&\frac{1}{4} \Big(-2 S_{\wt{(\wt{(il)}k)}l\wt{(k\wt{(lj)})}}+2 S_{\wt{(il)}l\wt{(lj)}}+S_{1l\wt{(\wt{(il)}k)}}+S_{2l\wt{(\wt{(il)}k)}}-S_{1l\wt{(il)}}-S_{2l\wt{(il)}}+S_{1l\wt{(k\wt{(lj)})}}\nn\\
&&~+S_{2l\wt{(k\wt{(lj)})}}-S_{1l\wt{(lj)}}-S_{2l\wt{(lj)}}\Big)f_3^0(\wt{(il)},k,\wt{(lj)}) A_{4}^{0}(\h{1},\wt{(\wt{(il)}k)},\h{2},\wt{(k\wt{(lj)})}) J_{2}^{(2)}(p_{\wt{(\wt{(il)}k)}},p_{\wt{(k\wt{(lj)})}})\Big\}.\nn\\
\end{eqnarray}
The large angle soft subtraction term is integrated analytically and added back to the real-virtual subtraction term.

\section{Real-virtual contribution}
\label{sec:RV}

The real-virtual matrix element is given by the interference of the one-loop amplitude with the tree,
\ba
\bs{A}_{5}^{1}(\{p\})&=&\la{\cal{A}}_{5}^{0}|{\cal{A}}_{5}^{1}\ra+\la{\cal{A}}_{5}^{1}|{\cal{A}}_{5}^{0}\ra.
\ea
In a particular colour ordered basis, the one-loop amplitude can be decomposed into the partial amplitudes~\cite{Bern:1993mq,Bern:1990ux},
\ba
\lefteqn{\bs{\cal{A}}_{5}^{1,\{a\}}(\{p\})=\la\bs{a}|{\cal{A}}_{5}^{1}(\{p\})\ra}\nn\\
&&\ \ \sum_{\sigma\in S_{5}/Z_{5}}N\ {\rm{Tr}}(a_{\sigma(1)},a_{\sigma(2)},a_{\sigma(3)},a_{\sigma(4)},a_{\sigma(5)})\ {\cal{A}}_{5,1}^{1}(\sigma(1),\sigma(2),\sigma(3),\sigma(4),\sigma(5))\nn\\
&+&\ \ \sum_{\rho\in S_{5}/Z_{4}}\ \ \ \ {\rm{Tr}}(a_{\rho(1)}){\rm{Tr}}(a_{\rho(2)},a_{\rho(3)},a_{\rho(4)},a_{\rho(5)})\ {\cal{A}}_{5,2}^{1}(\rho(1),\rho(2),\rho(3),\rho(4),\rho(5))\nn\\
&+&\sum_{\tau\in S_{5}/Z_{2}\times Z_{3}}\ \ {\rm{Tr}}(a_{\tau(1)}a_{\tau(2)}){\rm{Tr}}(a_{\tau(3)},a_{\tau(4)},a_{\tau(5)})\ {\cal{A}}_{5,3}^{1}(\tau(1),\tau(2),\tau(3),\tau(4),\tau(5))\nn\\\label{eq:m51decom}
\ea
where $\sigma$ is the set of orderings inequivalent under cyclic permutations.  $\rho$ is the set of orderings inequivalent under cyclic orderings of the subset of four elements, $\{\rho(2),\rho(3),\rho(4),\rho(5)\}$.  $\tau$ is the set of orderings inequivalent under cyclic permutations of the two subsets of orderings $\{\tau(1),\tau(2)\}$ and $\{\tau(3),\tau(4),\tau(5)\}$. The colour factors of the terms proportional to ${\cal{A}}_{5,2}^{1}$ in Eq.~\eqref{eq:m51decom} are identically zero.  

The sub-leading colour partial amplitude, ${\cal{A}}_{5,3}^{1}$ can be written in terms of the leading colour partial amplitude by using the decoupling identities~\cite{Bern:1990ux},
\ba
{\cal{A}}_{5,3}^{1}(1,2,3,4,5)&=&-{\cal{A}}_{5,2}^{1}(2,1,3,4,5)-{\cal{A}}_{5,2}^{1}(2,1,4,5,3)-{\cal{A}}_{5,2}^{1}(2,1,5,3,4)\label{eq:m53decoup}\\
{\cal{A}}_{5,2}^{1}(1,2,3,4,5)&=&-{\cal{A}}_{5,1}^{1}(1,2,3,4,5)-{\cal{A}}_{5,1}^{1}(1,3,4,5,2)\nn\\
&&-{\cal{A}}_{5,1}^{1}(1,4,5,2,3)-{\cal{A}}_{5,1}^{1}(1,5,2,3,4),\label{eq:m52decoup}
\ea
which leads to an expression for the real-virtual cross section in terms of interferences of leading colour one-loop partial amplitudes with tree-level amplitudes.

There are many ways to write the one-loop cross section due to the decoupling identities between partial amplitudes. It was shown in Sec.~\ref{sec:RR} that the double real cross section can be written in terms of the three independent interferences with no common neighbouring partons.  In the case of five gluon one-loop scattering there is only one independent ordering containing no common neighbouring partons such that,
\ba
\lefteqn{{\cal{SLC}}\Big(\bs{A}_{5}^{1}(\{p\})\Big)=12g^8 N^2(N^2-1)}\nn\\
&&\sum_{\sigma\in S_{5}/Z_{5}}2{\rm{Re}}\Big({\cal{A}}_{5}^{0,\dagger}(1,\sigma{(2)},\sigma{(3)},\sigma{(4)},\sigma{(5)}){\cal{A}}_{5,1}^{1}(1,\sigma{(4)},\sigma{(2)},\sigma{(5)},\sigma{(3)})\Big),
\ea 
and therefore the sub-leading colour one-loop five gluon cross section can be written in the optimal form,
\ba
\dsigma_{NNLO}^{RV}&=&{\cal N}_{LO}\left(\frac{\alpha_s}{2\pi}\right)^2\frac{\bar{C}(\epsilon)^2}{C(\epsilon)}\frac{12}{3!}\sum_{\sigma\in S_{5}/Z_{5}}
\int \frac{{\rm d}x_1}{x_1}\frac{{\rm d}x_2}{x_2}{\rm d}\Phi_{3}(p_3,\hdots,p_5;\bar{p}_1,\bar{p}_2)\;\nn\\
2{\rm{Re}}&\bigg\{&{\cal{A}}_{5}^{0,\dagger}(\h{\b{1}},\sigma(2),\sigma(3),\sigma(4),\sigma(5))\;{\cal{A}}_{5,1}^{1}(\h{\b{1}},\sigma(4),\sigma(2),\sigma(5),\sigma(3))\ \bigg\}\nn\\
&&\times \JET_{2}^{(3)}(p_i,p_j,p_k).\label{eq:rvcompact1}
\ea
This form for the sub-leading colour contribution to the five-gluon one-loop matrix element is equivalent to the expressions found in Eqs.~(9.12) and (9.13) in~\cite{Bern:1990ux} and greatly simplifies the construction of the real-virtual subtraction term. We have cross checked our numerical implementation of the sub-leading colour matrix element in Eq.~\eqref{eq:rvcompact1} against the numerical package {\tt NJET}~\cite{Badger:2012pg} and we find complete agreement between the two. By fixing the position of the second initial-state parton explicitly, the permutation sum reduces to a sum over final-state partons, 
\ba
\dsigma_{NNLO}^{RV}&=&{\cal N}_{LO}\left(\frac{\alpha_s}{2\pi}\right)^2\frac{\bar{C}(\epsilon)^2}{C(\epsilon)}\frac{24}{3!}\sum_{(i,j,k)\in P(3,4,5)}
\int \frac{{\rm d}x_1}{x_1}\frac{{\rm d}x_2}{x_2}{\rm d}\Phi_{3}(p_3,\hdots,p_5;\bar{p}_1,\bar{p}_2)\;\nn\\
2{\rm{Re}}&\bigg\{&{\cal{A}}_{5}^{0,\dagger}(\hb{1},\hb{2},i,j,k)\;{\cal{A}}_{5,1}^{1}(\h{\b{1}},j,\hb{2},k,i)-{\cal{A}}_{5}^{0,\dagger}(\hb{1},j,\hb{2},k,i)\;{\cal{A}}_{5,1}^{1}(\h{\b{1}},\hb{2},i,j,k)\ \bigg\}\nn\\
&&\times \JET_{2}^{(3)}(p_i,p_j,p_k).\label{eq:rvcompact2}
\ea
It should be noted that Eq.~\eqref{eq:rvcompact2} is simply a rearrangement of the sum in Eq.~\eqref{eq:rvcompact1} and is also free from collinear divergences.

The real-virtual subtraction term can be divided into three distinct contributions,
\ba
\dsigma_{NNLO}^{T}&=&\dsigma_{NNLO}^{T,a}+\dsigma_{NNLO}^{T,b}+\dsigma_{NNLO}^{T,c},
\ea
which will be explained in detail in the following sections.

\subsection{Explicit singularity subtraction}
\label{sec:dsigta}

The poles of a one-loop interference can be written in terms of integrated dipoles~\cite{Currie:2013vh},
\ba
\Poles\Big[2{\rm{Re}}\Big({\cal{A}}_{5}^{0,\dagger}(\sigma){\cal{A}}_{5,1}^{1}(\rho)\Big)\Big]&=&\sum_{{\rm{adj.pairs}}(i,j)\in\rho}-\frac{1}{2}\bs{J}_{2}^{(1)}(i,j)\ 2{\rm{Re}}\Big({\cal{A}}_{5}^{0,\dagger}(\sigma){\cal{A}}_{5}^{0}(\rho)\Big),\label{eq:m51poles}
\ea
where the choice of dipole (final-final, initial-final or initial-initial) depends of the kinematics of the radiators in the dipole.  Substituting Eq.~\eqref{eq:m51poles} into Eq.~\eqref{eq:rvcompact2} gives an expression for the poles of the full one-loop interference in terms of integrated dipoles,
\ba
\Poles\bigg(\dsigma_{NNLO}^{RV}\bigg)&=&{\cal N}_{LO}\left(\frac{\alpha_s}{2\pi}\right)^2\frac{\bar{C}(\epsilon)^2}{C(\epsilon)}\frac{12}{3!}\sum_{(i,j,k)\in P(3,4,5)}
\int \frac{{\rm d}x_1}{x_1}\frac{{\rm d}x_2}{x_2}{\rm d}\Phi_{3}(p_3,\hdots,p_5;\bar{p}_1,\bar{p}_2)\;\nn\\
2{\rm{Re}}\bigg\{&\bigg(&\bs{J}_{5}^{(1)}(\h{\b{1}},\hb{2},i,j,k)-\bs{J}_{5}^{(1)}(\h{\b{1}},j,\hb{2},k,i)\bigg)~{\cal{A}}_{5}^{0,\dagger}(\hb{1},\hb{2},i,j,k)\;{\cal{A}}_{5}^{0}(\h{\b{1}},j,\hb{2},k,i)\bigg\}\nn\\
&&\times\JET_{2}^{(3)}(p_i,p_j,p_k).\label{eq:rvpoles}
\ea
Eq.~\eqref{eq:rvpoles} can be written in terms of ten integrated dipoles using Eq.~\eqref{eq:jsum}. It should be noted that the mass factorization kernels in Eq.~\eqref{eq:rvpoles}  cancel and so the poles of the one-loop matrix element are given purely in terms of integrated antennae. These ten dipoles correspond to the ten antennae in the single unresolved subtraction term in Eq.~\eqref{eq:siga2}.

Explicitly carrying out the integration of the single unresolved subtraction term we find that,
\ba
\dsigma_{NNLO}^{T,a}&=&-\int_{1}\dsigma_{NNLO}^{S,a},\nn\\
&=&{\cal N}_{LO}\left(\frac{\alpha_s}{2\pi}\right)^2\frac{\bar{C}(\epsilon)^2}{C(\epsilon)}\frac{12}{3!}\sum_{(i,j,k)=P(3,4,5)}
\int \frac{{\rm d}x_1}{x_1}\frac{{\rm d}x_2}{x_2}{\rm d}\Phi_{3}(p_3,\hdots,p_5;\bar{p}_1,\bar{p}_2)\;\nn\\
\times2{\rm{Re}}&\bigg\{&\Big(\calF(s_{\b{1}\b{2}})+\frac{1}{2}\calF(s_{\b{2}i})+\frac{1}{3}\calF(s_{ij})+\frac{1}{3}\calF(s_{jk})+\frac{1}{2}\calF(s_{\b{1}k})\nn\\
&&-\frac{1}{2}\calF(s_{\b{1}j})-\frac{1}{2}\calF(s_{\b{2}j})-\frac{1}{2}\calF(s_{\b{2}k})-\frac{1}{3}\calF(s_{ik})-\frac{1}{2}\calF(s_{\b{1}i})\bigg) \nn\\
&\times&~{\cal{A}}_{5}^{0,\dagger}(\hb{1},\hb{2},i,j,k)\;{\cal{A}}_{5}^{0}(\h{\b{1}},j,\hb{2},k,i)\bigg\}\JET_{2}^{(3)}(p_i,p_j,p_k).\label{eq:dsigta}
\ea
Eq.~\eqref{eq:dsigta} clearly matches the form for the poles of the 1-loop matrix element in Eq.~\eqref{eq:rvpoles} and so that,
\ba
\Poles\bigg(\dsigma_{NNLO}^{RV}-\dsigma_{NNLO}^{T,a}\bigg)&=&0.
\ea

\subsection{Implicit singularity subtraction}

In single unresolved regions of phase space the jet function allows the real-virtual matrix element to develop implicit divergences.  In order to be able to integrate the real-virtual cross section numerically, a single unresolved subtraction term is constructed to remove any implicit singularities of the real-virtual cross section.

The form of the cross section in Eq.~\eqref{eq:rvcompact1} makes it particularly clear that the total cross section contains no divergent collinear limits at sub-leading colour; this leaves only soft limits to consider. In the single soft limit, the one-loop colour ordered amplitudes factorize in the following way,
\ba
{\cal{A}}_{5,1}^{1}(\cdots,a,i,b,\cdots)&\stackrel{i\to0}{\longrightarrow}&\S^{0}(p_{a},p_{i},p_{b}){\cal{A}}_{4,1}^{1}(\cdots,a,b,\cdots)\nn\\
&+&\S^{1}(p_{a},p_{i},p_{b}){\cal{A}}_{4}^{0}(\cdots,a,b,\cdots),\label{eq:m51softfac}
\ea
where the colour stripped one-loop soft function\cite{Bern:1998sc,Bern:1999ry,Catani:2000pi} can be written in the form,
\ba
\S^{1}(p_{a},p_{i},p_{b})&=&-\S^{0}(p_{a},p_{i},p_{b})\cdot\S^{{\rm{sing}}}(a,i,b)\label{eq:s1},
\ea
and the singular function, ${\cal{S}}^{{\rm{sing}}}$, is given by,
\ba
{\cal{S}}^{{\rm{sing}}}(a,i,b)&=&\b{C}(\e)\frac{1}{\e^2}\bigg(-\frac{\mu^{2}s_{ab}}{s_{ai}s_{ib}}\bigg)^{\e}.
\ea
Substituting Eqs.~\eqref{eq:m51softfac} and~\eqref{eq:s1} into the sub-leading colour contribution shown in Eq.~\eqref{eq:rvcompact2}, yields an expression where each term containing a one-loop soft current is purely imaginary and so does not contribute to the matrix element. This shows that the one-loop soft gluon current does not contribute to the single unresolved limit of the sub-leading colour one-loop matrix element and so no one-loop antennae are required in the real-virtual subtraction term. Similarly in the colour space approach, the one-loop soft gluon current is proportional to the sandwich $\la{\cal{A}}_{4}^{0}|\bt_{i}\cdot\bt_{j}|{\cal{A}}_{4}^{0}\ra$, which vanishes at sub-leading colour, as stated in Eq.~\eqref{eq:slczero}.

Promoting the eikonal factors of the soft limit to three-parton antennae leads us to the following subtraction term for a generic one-loop interference,
\ba
\lefteqn{{\cal{A}}_{5}^{0,\dagger}(\cdots,a,i,b,\cdots){\cal{A}}_{5}^{1}(\cdots,c,i,d,\cdots)\stackrel{i\to0}{\approx}}\nn\\
&+&X_{3}^{0}(a,i,c)\ {\cal{A}}_{4}^{0,\dagger}(\cdots,\wt{(ai)},{b},\cdots){\cal{A}}_{4,1}^{1}(\cdots,\wt{(ic)},{d},\cdots)\nn\\
&+&X_{3}^{0}(b,i,d)\ {\cal{A}}_{4}^{0,\dagger}(\cdots,{a},\wt{(bi)},\cdots){\cal{A}}_{4,1}^{1}(\cdots,{c},\wt{(id)},\cdots)\nn\\
&-&X_{3}^{0}(a,i,d)\ {\cal{A}}_{4}^{0,\dagger}(\cdots,\wt{(ai)},{b},\cdots){\cal{A}}_{4,1}^{1}(\cdots,{c},\wt{(id)},\cdots)\nn\\
&-&X_{3}^{0}(b,i,c)\ {\cal{A}}_{4}^{0,\dagger}(\cdots,{a},\wt{(bi)},\cdots){\cal{A}}_{4,1}^{1}(\cdots,\wt{(ic)},{d},\cdots),\label{eq:1loopfact}
\ea
where once again, the explicit form of $X_{3}^{0}$ depends on the kinematics of the hard radiators. Applying Eq.~\eqref{eq:1loopfact} to the real-virtual cross section in Eq.~\eqref{eq:rvcompact2} and simplifying the result yields the single unresolved subtraction term,
\begin{eqnarray}
&&\dsigma_{NNLO}^{T,b}={\cal N}_{LO}\left(\frac{\alpha_s}{2\pi}\right)^2\frac{\bar{C}(\epsilon)^2}{C(\epsilon)}\frac{12}{3!}\sum_{(i,j,k)\in P(3,4,5)}
\int \frac{{\rm d}x_1}{x_1}\frac{{\rm d}x_2}{x_2}{\rm d}\Phi_{3}(p_3,\hdots,p_5;\bar{p}_1,\bar{p}_2)\;2{\rm{Re}}\Big\{\nn\\
&&+2F_3^0(\hb{1},k,\hb{2}) \Big[{\cal{A}}_4^{0\dagger}(\hbb{1},\hbb{2},\t{i},\t{j})\Big( {\cal{A}}_4^1(\hbb{1},\t{i},\hbb{2},\t{j})\delta_1 \delta_{2}+\frac{1}{2}\bs{J}_{4}^{(1)}(\hbb{1},\t{i},\hbb{2},\t{j}){\cal{A}}_{4}^{0}(\hbb{1},\t{i},\hbb{2},\t{j})\Big)\nn\\
&&\hspace{2.0cm}-\, {\cal{A}}_4^{0\dagger}(\hbb{1},\t{i},\hbb{2},\t{j}) \Big({\cal{A}}_4^1(\hbb{1},\hbb{2},\t{i},\t{j})\delta_1 \delta_{2}+\frac{1}{2}\bs{J}_{4}^{(1)}(\hbb{1},\hbb{2},\t{i},\t{j}){\cal{A}}_{4}^{0}(\hbb{1},\hbb{2},\t{i},\t{j})\Big)\Big] J_{2}^{(2)}(p_{\t{i}},p_{\t{j}})\nn\\
&&+2f_3^0(\hb{1},k,i)\times\nn\\
&&\ \  \Big[ {\cal{A}}_4^{0\dagger}(\hbb{1},\hb{2},j,\wt{(ki)})\Big( {\cal{A}}_4^1(\hbb{1},\hb{2},\wt{(ki)},j)\delta_1 \delta_{2}+\frac{1}{2}\bs{J}_{4}^{(1)}(\hbb{1},\hb{2},\wt{(ki)},j){\cal{A}}_{4}^{0}(\hbb{1},\hb{2},\wt{(ki)},j)\Big)\nn\\
&&\ \ - {\cal{A}}_4^{0\dagger}(\hbb{1},\hb{2},\wt{(ki)},j)\Big( {\cal{A}}_4^1(\hbb{1},\hb{2},j,\wt{(ki)})\delta_1 \delta_{2}+\frac{1}{2}\bs{J}_{4}^{(1)}(\hbb{1},\hb{2},j,\wt{(ki)}){\cal{A}}_{4}^{0}(\hbb{1},\hb{2},j,\wt{(ki)})\Big)\nn\\
&&\ \ + {\cal{A}}_4^{0\dagger}(\hbb{1},\wt{(ki)},\hb{2},j)\Big( {\cal{A}}_4^1(\hbb{1},\hb{2},\wt{(ki)},j)\delta_1 \delta_{2}+\frac{1}{2}\bs{J}_{4}^{(1)}(\hbb{1},\hb{2},\wt{(ki)},j){\cal{A}}_{4}^{0}(\hbb{1},\hb{2},\wt{(ki)},j)\Big)\nn\\
&&\ \ - {\cal{A}}_4^{0\dagger}(\hbb{1},\hb{2},\wt{(ki)},j)\Big( {\cal{A}}_4^1(\hbb{1},\wt{(ki)},\hb{2},j)\delta_1 \delta_{2}+\frac{1}{2}\bs{J}_{4}^{(1)}(\hbb{1},\wt{(ki)},\hb{2},j){\cal{A}}_{4}^{0}(\hbb{1},\wt{(ki)},\hb{2},j)\Big)\Big]J_{2}^{(2)}(p_{\wt{(ki)}},p_{j})\nn\\
&&+2f_3^0(\h{2},k,i) \times\nn\\
&&\ \ \Big[{\cal{A}}_4^{0\dagger}(\hb{1},\hbb{2},\wt{(ki)},j)\Big( {\cal{A}}_4^1(\hb{1},\hbb{2},j,\wt{(ki)})\delta_1 \delta_{2}+\frac{1}{2}\bs{J}_{4}^{(1)}(\hb{1},\hbb{2},j,\wt{(ki)}){\cal{A}}_{4}^{0}(\hb{1},\hbb{2},j,\wt{(ki)})\Big)\nn\\
&&\ \ - {\cal{A}}_4^{0\dagger}(\hb{1},\hbb{2},j,\wt{(ki)})\Big( {\cal{A}}_4^1(\hb{1},\hbb{2},\wt{(ki)},j)\delta_1 \delta_{2}+\frac{1}{2}\bs{J}_{4}^{(1)}(\hb{1},\hbb{2},\wt{(ki)},j){\cal{A}}_{4}^{0}(\hb{1},\hbb{2},\wt{(ki)},j)\Big)\nn\\
&&\ \ + {\cal{A}}_4^{0\dagger}(\hb{1},\wt{(ki)},\hbb{2},j)\Big( {\cal{A}}_4^1(\hb{1},\hbb{2},j,\wt{(ki)})\delta_1 \delta_{2}+\frac{1}{2}\bs{J}_{4}^{(1)}(\hb{1},\hbb{2},j,\wt{(ki)}){\cal{A}}_{4}^{0}(\hb{1},\hbb{2},j,\wt{(ki)})\Big)\nn\\
&&\ \ - {\cal{A}}_4^{0\dagger}(\hb{1},\hbb{2},j,\wt{(ki)})\Big( {\cal{A}}_4^1(\hb{1},\wt{(ki)},\hbb{2},j)\delta_1 \delta_{2}+\frac{1}{2}\bs{J}_{4}^{(1)}(\hb{1},\wt{(ki)},\hbb{2},j){\cal{A}}_{4}^{0}(\hb{1},\wt{(ki)},\hbb{2},j)\Big)\Big] J_{2}^{(2)}(p_j,p_{\wt{(ki)}})\nn\\
&&+f_3^0(i,k,j)\times\nn\\
&&\ \ \Big[\;{\cal{A}}_4^{0\dagger}(\hb{1},\hb{2},\wt{(ik)},\wt{(kj)})\Big( {\cal{A}}_4^1(\hb{1},\wt{(ik)},\hb{2},\wt{(kj)})\delta_1 \delta_{2}+\frac{1}{2}\bs{J}_{4}^{(1)}(\hb{1},\wt{(ik)},\hb{2},\wt{(kj)}){\cal{A}}_{4}^{0}(\hb{1},\wt{(ik)},\hb{2},\wt{(kj)})\Big)\nn\\
&&\ \ -{\cal{A}}_4^{0\dagger}(\hb{1},\wt{(ik)},\hb{2},\wt{(kj)})\Big( {\cal{A}}_4^1(\hb{1},\hb{2},\wt{(ik)},\wt{(kj)})\delta_1 \delta_{2}+\frac{1}{2}\bs{J}_{4}^{(1)}(\hb{1},\hb{2},\wt{(ik)},\wt{(kj)}){\cal{A}}_{4}^{0}(\hb{1},\hb{2},\wt{(ik)},\wt{(kj)})\Big)\nn\\
&&\ \ +{\cal{A}}_4^{0\dagger}(\hb{1},\hb{2},\wt{(kj)},\wt{(ik)})\Big( {\cal{A}}_4^1(\hb{1},\wt{(ik)},\hb{2},\wt{(kj)})\delta_1 \delta_{2}+\frac{1}{2}\bs{J}_{4}^{(1)}(\hb{1},\wt{(ik)},\hb{2},\wt{(kj)}){\cal{A}}_{4}^{0}(\hb{1},\wt{(ik)},\hb{2},\wt{(kj)})\Big)\nn\\
&&\ \  -{\cal{A}}_4^{0\dagger}(\hb{1},\wt{(ik)},\hb{2},\wt{(kj)})\Big( {\cal{A}}_4^1(\hb{1},\hb{2},\wt{(kj)},\wt{(ik)})\delta_1 \delta_{2}+\frac{1}{2}\bs{J}_{4}^{(1)}(\hb{1},\hb{2},\wt{(kj)},\wt{(ik)}){\cal{A}}_{4}^{0}(\hb{1},\hb{2},\wt{(kj)},\wt{(ik)})\Big)\Big]\nn\\
&&\ \ \times J_{2}^{(2)}(p_{\wt{(ik)}},p_{\wt{(kj)}})\Big\},\label{eq:dstb}
\end{eqnarray}
where $\delta_{1,2}=\delta(1-x_{1,2})$. The integrated antenna strings, $\bs{J}_{4}^{(1)}$, in Eq.~\eqref{eq:dstb} are introduced to remove the explicit IR poles of the reduced four gluon one-loop amplitudes.  Once again, any mass factorization kernels in the integrated dipoles cancel.  For ease of exposition in later sections, we will refer to those terms in Eq.~\eqref{eq:dstb} that are proportional to the one-loop amplitudes as $\dsigma_{NNLO}^{T,b_{1}}$ and those proportional to the integrated dipoles as $\dsigma_{NNLO}^{T,b_{2}}$. Both of these terms have to be integrated analytically and added to the double virtual subtraction term.

\subsection{Spurious singularity subtraction}
\label{sec:dstc}

There are additional double real subtraction terms,  $\dsigma_{NNLO}^{S,b}$, $\dsigma_{NNLO}^{S,c}$ and $\dsigma_{NNLO}^{S,e}$, that are added to the real-virtual cross section after analytic integration. It is useful to consider the spurious singularity subtraction term as a sum of two contributions,
\ba
\dsigma_{NNLO}^{T,c}&=&\dsigma_{NNLO}^{T,c_{1}}+\dsigma_{NNLO}^{T,c_{2}},
\ea
where $\dsigma_{NNLO}^{T,c_{1}}$ consists of the terms inherited directly from the double real subtraction terms after analytic integration, 
\ba
\dsigma_{NNLO}^{T,c_{1}}&=&-\int_{1}\Big[\dsigma_{NNLO}^{S,b}+\dsigma_{NNLO}^{S,c}+\dsigma_{NNLO}^{S,e}\Big].\label{eq:dstc1def}
\ea
The subtraction term in Eq.~\eqref{eq:dstc1def} produces explicit poles and implicit divergences in the real-virtual contribution.  Since all explicit and implicit singularities in the real-virtual cross section have already been removed by $\dsigma_{NNLO}^{T,a}$ and $\dsigma_{NNLO}^{T,b}$, the singularities introduced by Eq.~\eqref{eq:dstc1def} must be explicitly cancelled by an additional  subtraction term, $\dsigma_{NNLO}^{T,c_2}$. 

Following~\cite{GehrmannDeRidder:2011aa} we find that the spurious singularity subtraction term is given by,
\ba
\lefteqn{\dsigma_{NNLO}^{T,c}={\cal N}_{LO}\left(\frac{\alpha_s}{2\pi}\right)^2\frac{\bar{C}(\epsilon)^2}{C(\epsilon)}\frac{12}{3!}\sum_{(i,j,k)\in P(3,4,5)}
\int \frac{{\rm d}x_1}{x_1}\frac{{\rm d}x_2}{x_2}{\rm d}\Phi_{3}(p_3,\hdots,p_5;\bar{p}_1,\bar{p}_2)\Big\{}\nn\\
+\frac{1}{2}&\Big[&\calF(s_{\b{1}i})-\calF(s_{\bb{1}(ki)})+\frac{1}{2}\calF(s_{\bb{1}j})-\frac{1}{2}\calF(s_{\b{1}j})+\frac{1}{2}\calF(s_{\b{2}(ki)})-\frac{1}{2}\calF(s_{\b{2}i})+\calF(s_{\bb{1}\b{2}})\nn\\
&-&\calF(s_{\b{1}\b{2}})+\frac{1}{3}\calF(s_{(ki)j})-\frac{1}{3}\calF(s_{ij})+2{\cal{S}}(s_{\b{1}\wt{(ki)}},s_{ij})-2{\cal{S}}(s_{\b{1}i},s_{ij})+{\cal{S}}(s_{ij},s_{ij})\nn\\
&-&{\cal{S}}(s_{\wt{(ki)j}},s_{ij})+{\cal{S}}(s_{\b{2}i},s_{ij})-{\cal{S}}(s_{\b{2}\wt{(ki)}},s_{ij})\Big]~ f_{3}^{0}(\hb{1},k,i)~A_{4}^{0}(\hbb{1},\hb{2},\wt{(ki)},j)~J_{2}^{(2)}(p_{j},p_{\wt{(ki)}})\nn\\\nn\\
+\frac{1}{2}&\Big[&\calF(s_{\b{2}j})-\calF(s_{\bb{2}(kj)})+\frac{1}{2}\calF(s_{\b{1}(kj)})-\frac{1}{2}\calF(s_{\b{1}j})+\frac{1}{2}\calF(s_{\bb{2}i})-\frac{1}{2}\calF(s_{\b{2}i})+\calF(s_{\b{1}\bb{2}})\nn\\
&-&\calF(s_{\b{1}\b{2}})+\frac{1}{3}\calF(s_{i(kj)})-\frac{1}{3}\calF(s_{ij})+2{\cal{S}}(s_{\b{2}\wt{(kj)}},s_{ij})-2{\cal{S}}(s_{\b{2}j},s_{ij})+{\cal{S}}(s_{ij},s_{ij})\nn\\
&-&{\cal{S}}(s_{\wt{i(kj)}},s_{ij})+{\cal{S}}(s_{\b{1}j},s_{ij})-{\cal{S}}(s_{\b{1}\wt{(kj)}},s_{ij})\Big]~f_{3}^{0}(\hb{2},k,j)~A_{4}^{0}(\hb{1},\hbb{2},i,\wt{(kj)})~J_{2}^{(2)}(p_{i},p_{\wt{(kj)}})\nn\\\nn\\
-\frac{1}{2}&\Big[&\frac{1}{2}\calF(s_{\b{1}i})-\frac{1}{2}\calF(s_{\bb{1}\tilde{i}})+\frac{1}{2}\calF(s_{\b{2}j})-\frac{1}{2}\calF(s_{\bb{2}\tilde{j}})+\frac{1}{2}\calF(s_{\bb{1}\tilde{j}})-\frac{1}{2}\calF(s_{\b{1}j})+\frac{1}{2}\calF(s_{\bb{2}\tilde{i}})\nn\\
&-&\frac{1}{2}\calF(s_{\b{2}i})+{\cal{S}}(s_{\bb{1}\tilde{i}},s_{\tilde{i}\tilde{j}})-{\cal{S}}(s_{\b{1}i},s_{ij})+{\cal{S}}(s_{\bb{2}\tilde{j}},s_{\tilde{i}\tilde{j}})-{\cal{S}}(s_{\b{2}j},s_{ij})-{\cal{S}}(s_{\bb{1}\tilde{j}},s_{\tilde{i}\tilde{j}})\nn\\
&+&{\cal{S}}(s_{\b{1}j},s_{ij})-{\cal{S}}(s_{\bb{2}\tilde{i}},s_{\tilde{i}\tilde{j}})+{\cal{S}}(s_{\b{2}i},s_{ij})\Big]~F_{3}^{0}(\hb{1},k,\hb{2})~A_{4}^{0}(\hbb{1},\hbb{2},\tilde{i},\tilde{j})~J_{2}^{(2)}(p_{\tilde{i}},p_{\tilde{j}})\nn\\\nn\\
-\frac{1}{4}&\Big[&\frac{1}{2}\calF(s_{\b{1}i})-\frac{1}{2}\calF(s_{\b{1}(ik)})+\frac{1}{2}\calF(s_{\b{2}j})-\frac{1}{2}\calF(s_{\b{2}(kj)})+\frac{1}{2}\calF(s_{\b{1}(kj)})-\frac{1}{2}\calF(s_{\b{1}j})\nn\\
&+&\frac{1}{2}\calF(s_{\b{2}(ik)})-\frac{1}{2}\calF(s_{\b{2}i})+{\cal{S}}(s_{\b{1}\wt{(ik)}},s_{ij})-{\cal{S}}(s_{\b{1}i},s_{ij})+{\cal{S}}(s_{\b{2}\wt{(kj)}},s_{ij})\nn\\
&-&{\cal{S}}(s_{\b{2}j},s_{ij})+{\cal{S}}(s_{\b{1}j},s_{ij})-{\cal{S}}(s_{\b{1}\wt{(kj)}},s_{ij})+{\cal{S}}(s_{\b{2}i},s_{ij})-{\cal{S}}(s_{\b{2}\wt{(ki)}},s_{ij})\Big]\nn\\
&&\times~f_{3}^{0}(i,k,j)~A_{4}^{0}(\hb{1},\hb{2},\wt{(ik)},\wt{(kj)})~J_{2}^{(2)}(p_{\wt{(ik)}},p_{\wt{(kj)}})\nn\\\nn\\
+\frac{1}{4}&\Big[&\frac{1}{2}\calF(s_{\b{1}i})-\frac{1}{2}\calF(s_{\b{1}(ik)})+\frac{1}{2}\calF(s_{\b{2}j})-\frac{1}{2}\calF(s_{\b{2}(kj)})+\frac{1}{2}\calF(s_{\b{1}(kj)})-\frac{1}{2}\calF(s_{\b{1}j})\nn\\
&+&\frac{1}{2}\calF(s_{\b{2}(ik)})-\frac{1}{2}\calF(s_{\b{2}i})+{\cal{S}}(s_{\b{1}\wt{(ik)}},s_{ij})-{\cal{S}}(s_{\b{1}{i}},s_{ij})+{\cal{S}}(s_{\b{2}\wt{(kj)}},s_{ij})\nn\\
&-&{\cal{S}}(s_{\b{2}{j}},s_{ij})+{\cal{S}}(s_{\b{1}{j}},s_{ij})-{\cal{S}}(s_{\b{1}\wt{(kj)}},s_{ij})+{\cal{S}}(s_{\b{2}{i}},s_{ij})-{\cal{S}}(s_{\b{2}\wt{(ik)}},s_{ij})\Big]\nn\\
&&\times~f_{3}^{0}(i,k,j)~A_{4}^{0}(\hb{1},\hb{2},\wt{(kj)},\wt{(ik)})~J_{2}^{(2)}(p_{\wt{(ik)}},p_{\wt{(kj)}})\nn\\\nn\\
+\frac{1}{2}&\Big[&\calF(s_{\b{1}\b{2}})-\calF(s_{\bb{1}\b{2}})+\frac{1}{3}\calF(s_{ij})-\frac{1}{3}\calF(s_{(ki)j})-\frac{1}{2}\calF(s_{\b{1}j})+\frac{1}{2}\calF(s_{\bb{1}j})\nn\\
&-&\frac{1}{2}\calF(s_{\b{2}i})+\frac{1}{2}\calF(s_{\b{2}\wt{(ki)}})+{\cal{S}}(s_{\wt{(ki)}j},s_{ij})-{\cal{S}}(s_{ij},s_{ij})+{\cal{S}}(s_{\b{2}i},s_{ij})-{\cal{S}}(s_{\b{2}\wt{(ki)}},s_{ij})\Big]\nn\\
&&\times~f_{3}^{0}(\hb{1},k,i)~A_{4}^{0}(\hbb{1},\hb{2},j,\wt{(ki)})~J_{2}^{(2)}(p_{j},p_{\wt{(ki)}})\nn\\\nn\\
+\frac{1}{2}&\Big[&\calF(s_{\b{1}\b{2}})-\calF(s_{\b{1}\bb{2}})+\frac{1}{3}\calF(s_{ij})-\frac{1}{3}\calF(s_{i(kj)})-\frac{1}{2}\calF(s_{\b{1}j})+\frac{1}{2}\calF(s_{\b{1}(kj)})\nn\\
&-&\frac{1}{2}\calF(s_{\b{2}i})+\frac{1}{2}\calF(s_{\bb{2}i})+{\cal{S}}(s_{i\wt{(kj)}},s_{ij})-{\cal{S}}(s_{ij},s_{ij})+{\cal{S}}(s_{\b{1}j},s_{ij})-{\cal{S}}(s_{\b{1}\wt{(kj)}},s_{ij})\Big]\nn\\
&&\times~f_{3}^{0}(\hb{2},k,j)~A_{4}^{0}(\hb{1},\hbb{2},\wt{(kj)},i)~J_{2}^{(2)}(p_{i},p_{\wt{(kj)}})\nn\\\nn\\
+\frac{1}{4}&\Big[&2\calF(s_{\b{1}\b{2}})-2\calF(s_{\bb{1}\bb{2}})-\frac{1}{2}\calF(s_{\b{1}{j}})+\frac{1}{2}\calF(s_{\bb{1}\tilde{j}})-\frac{1}{2}\calF(s_{\b{2}{i}})+\frac{1}{2}\calF(s_{\bb{2}\tilde{i}})-\frac{1}{2}\calF(s_{\b{1}{i}})\nn\\
&+&\frac{1}{2}\calF(s_{\bb{1}\tilde{i}})-\frac{1}{2}\calF(s_{\b{2}{j}})+\frac{1}{2}\calF(s_{\bb{2}\tilde{j}})+2{\cal{S}}(s_{\bb{1}\bb{2}},s_{\tilde{i}\tilde{j}})-2{\cal{S}}(s_{\b{1}\b{2}},s_{ij})-{\cal{S}}(s_{\bb{1}\tilde{j}},s_{\tilde{i}\tilde{j}})+{\cal{S}}(s_{\b{1}{j}},s_{ij})\nn\\
&-&{\cal{S}}(s_{\bb{2}\tilde{i}},s_{\tilde{i}\tilde{j}})+{\cal{S}}(s_{\b{2}{i}},s_{ij})-{\cal{S}}(s_{\bb{1}\tilde{i}},s_{\tilde{i}\tilde{j}})+{\cal{S}}(s_{\b{1}{i}},s_{ij})-{\cal{S}}(s_{\bb{2}\tilde{j}},s_{\tilde{i}\tilde{j}})+{\cal{S}}(s_{\b{2}{j}},s_{ij})\Big]\nn\\
&&\times~F_{3}^{0}(\hb{1},k,\hb{2})~A_{4}^{0}(\hbb{1},\tilde{i},\hbb{2},\tilde{j})~J_{2}^{(2)}(p_{\tilde{i}},p_{\tilde{j}})\nn\\\nn\\
+\frac{1}{4}&\Big[&\frac{2}{3}\calF(s_{ij})-\frac{2}{3}\calF(s_{(ik)(kj)})-\frac{1}{2}\calF(s_{\b{1}j})+\frac{1}{2}\calF(s_{\b{1}(kj)})-\frac{1}{2}\calF(s_{\b{2}i})+\frac{1}{2}\calF(s_{\b{2}(ik)})\nn\\
&-&\frac{1}{2}\calF(s_{\b{1}i})+\frac{1}{2}\calF(s_{\b{1}(ik)})-\frac{1}{2}\calF(s_{\b{2}j})+\frac{1}{2}\calF(s_{\b{2}(kj)})+2{\cal{S}}(s_{\wt{(ik)}\wt{(kj)}},s_{ij})-2{\cal{S}}(s_{ij},s_{ij})\nn\\
&-&{\cal{S}}(s_{\b{1}\wt{(kj)}},s_{ij})+{\cal{S}}(s_{\b{1}j},s_{ij})-{\cal{S}}(s_{\b{2}\wt{(ik)}},s_{ij})+{\cal{S}}(s_{\b{2}i},s_{ij})-{\cal{S}}(s_{\b{1}\wt{(ik)}},s_{ij})+{\cal{S}}(s_{\b{1}i},s_{ij})\nn\\
&-&{\cal{S}}(s_{\b{2}\wt{(kj)}},s_{ij})+{\cal{S}}(s_{\b{2}j},s_{ij})\Big]~f_{3}^{0}(i,k,j)~A_{4}^{0}(\hb{1},\wt{(ik)},\hb{2},\wt{(kj)})~J_{2}^{(2)}(p_{\wt{(ik)}},p_{\wt{(kj)}})\nn\\\nn\\
-\frac{1}{2}&\Big[&\calF(s_{\b{1}\b{2}})-\calF(s_{\bb{1}\b{2}})+\frac{1}{3}\calF(s_{ij})-\frac{1}{3}\calF(s_{(ki)j})-\frac{1}{2}\calF(s_{\b{1}j})+\frac{1}{2}\calF(s_{\bb{1}j})\nn\\
&-&\frac{1}{2}\calF(s_{\b{2}i})+\frac{1}{2}\calF(s_{\b{2}(ki)})+{\cal{S}}(s_{\wt{(ki)}j},s_{ij})-{\cal{S}}(s_{ij},s_{ij})+{\cal{S}}(s_{\b{2}i},s_{ij})-{\cal{S}}(s_{\b{2}\wt{(ki)}},s_{ij})\Big]\nn\\
&&\times~f_{3}^{0}(\hb{1},k,i)~A_{4}^{0}(\hbb{1},\wt{(ki)},\hb{2},j)~J_{2}^{(2)}(p_{j},p_{\wt{(ki)}})\nn\\\nn\\
-\frac{1}{2}&\Big[&\calF(s_{\b{1}\b{2}})-\calF(s_{\b{1}\bb{2}})+\frac{1}{3}\calF(s_{ij})-\frac{1}{3}\calF(s_{i(kj)})-\frac{1}{2}\calF(s_{\b{1}j})+\frac{1}{2}\calF(s_{\b{1}(kj)})\nn\\
&-&\frac{1}{2}\calF(s_{\b{2}i})+\frac{1}{2}\calF(s_{\bb{2}i})+{\cal{S}}(s_{i\wt{(kj)}},s_{ij})-{\cal{S}}(s_{ij},s_{ij})+{\cal{S}}(s_{\b{1}j},s_{ij})-{\cal{S}}(s_{\b{1}\wt{(kj)}},s_{ij})\Big]\nn\\
&&\times~f_{3}^{0}(\hb{2},k,j)A_{4}^{0}(\hb{1},i,\hbb{2},\wt{(kj)})J_{2}^{(2)}(p_{i},p_{\wt{(kj)}})\Big\}.\label{eq:dstc}
\ea
In Eq.~\eqref{eq:dstc}, the terms corresponding to $\dsigma_{NNLO}^{T,c_{2}}$ are those proportional to integrated antennae with mapped momentum arguments.  These terms are integrated analytically over the remaining unresolved phase space and added back into the double virtual cross section.  All other terms constitute $\dsigma_{NNLO}^{T,c_{1}}$ and terminate in the real-virtual cross section.

At this point we have fully constructed the subtraction terms which are used to remove all explicit IR poles and implicit IR divergences from $\dsigma_{NNLO}^{RV}$.  The pattern of singularity cancellation can be summarised as follows:
\begin{itemize}
\item The explicit poles of $\dsigma_{NNLO}^{RV}$ are cancelled by $\dsigma_{NNLO}^{T,a}$.
\item The implicit divergences of $\dsigma_{NNLO}^{RV}$ are removed by $\dsigma_{NNLO}^{T,b_{1}}$.
\item The explicit poles of $\dsigma_{NNLO}^{T,b_{1}}$ are cancelled by $\dsigma_{NNLO}^{T,b_{2}}$.
\item The implicit divergences of $\dsigma_{NNLO}^{T,b_{2}}$ cancel against those of $\dsigma_{NNLO}^{T,a}$.
\item $\dsigma_{NNLO}^{T,c}$ is free from poles in $\e$ and finite in all unresolved limits.
\end{itemize}

\section{Double virtual contribution}
\label{sec:VV}

The poles of the full colour double virtual matrix element can be expressed in terms of single and double unresolved integrated dipoles according to the formula~\cite{Currie:2013vh},
\ba
\lefteqn{\Poles\bigg(\dsigma_{NNLO}^{VV}\bigg)=-{\cal N}_{LO}\left(\frac{\alpha_s}{2\pi}\right)^2\bar{C}(\epsilon)^2
\int \frac{{\rm d}z_1}{z_1}\frac{{\rm d}z_2}{z_2}~{\rm d}\Phi_{3}(p_3,p_4;\bar{p}_1,\bar{p}_2)}\nn\\
2&\bigg\{&\sum_{(i,j)}\bs{J}_{2}^{(1)}(i,j)\ 2{\rm{Re}}\la{\cal A}_{4}^{0}|\bt_{i}\cdot\bt_{j}|{\cal A}_{4}^{1}\ra\nn\\
&-&\sum_{(i,j)}\sum_{(k,l)}\big[\bs{J}_{2}^{(1)}(i,j)\otimes\bs{J}_{2}^{(1)}(k,l)\big]\ \la{\cal{A}}_{4}^{0}|(\bt_{i}\cdot\bt_{j})(\bt_{k}\cdot\bt_{l})|{\cal A}_{4}^{0}\ra\nn\\
&+&\sum_{(i,j)}N\bs{J}_{2}^{(2)}(i,j)\ \la{\cal A}_{4}^{0}|\bt_{i}\cdot\bt_{j}|{\cal A}_{4}^{0}\ra\bigg\}~J_{2}^{(2)}(p_{3},p_{4}).\label{eq:vvpoles}
\ea
This expression has been confirmed by comparing to the analytic formulae for the two-loop interferences in~\cite{Glover:2001af,Glover:2001rd}.  The task of this section is then to demonstrate that the double virtual subtraction term matches this form for the double virtual pole structure.  The fact that the poles of the two-loop matrix element are written in terms of integrated antennae makes this demonstration particularly transparent.

The sub-leading colour double virtual subtraction term has three contributions,
\ba
\dsigma_{NNLO}^{U}&=&\dsigma_{NNLO}^{U,a}+\dsigma_{NNLO}^{U,b}+\dsigma_{NNLO}^{U,c}.
\ea
The subtraction term $\dsigma_{NNLO}^{U,a}$ corresponds to the first line of Eq.~\eqref{eq:vvpoles} containing the sandwiches involving one-loop amplitudes, $\dsigma_{NNLO}^{U,b}$ corresponds to the second line containing double colour charge insertions to tree-level sandwiches. The last term, $\dsigma_{NNLO}^{U,c}$, corresponds to the final line of Eq.~\eqref{eq:vvpoles} containing the double unresolved integrated dipole, $\bs{J}_{2}^{(2)}$.  This term is proportional to the sandwich $\la{\cal A}_{4}^{0}|\bt_{i}\cdot\bt_{j}|{\cal A}_{4}^{0}\ra$, which has no contribution at sub-leading colour according to Eq.~\eqref{eq:slczero}, and so,
\ba
\dsigma_{NNLO}^{U,c}&=&0.
\ea
The double unresolved integrated dipole, $\bs{J}_{2}^{(2)}$, is the only contribution that contains the integrated four-parton, ${\cal{X}}_{4}^{0}$, and one-loop, ${\cal{X}}_{3}^{1}$, antennae. Its absence from the sub-leading colour double virtual subtraction term implies that neither of these types of antennae are present, in unintegrated form, in the double real or real-virtual subtraction terms respectively. This is indeed what was found when explicitly constructing the double real and real-virtual subtraction terms in Secs.~\ref{sec:RR} and \ref{sec:RV}.

\subsection*{Single operator insertions into one-loop sandwiches}

The two-loop contribution contains a subset of poles which can be written in terms of colour charge insertions to the one-loop interferences of the type,
\ba
\bs{J}_{2}^{(1)}(i,j)\ 2{\rm{Re}}\la{\cal{A}}_{4}^{0}|\bt_{i}\cdot\bt_{j}|{\cal{A}}_{4}^{1}\ra.\label{eq:m41sandwich}
\ea
To evaluate these sandwiches explicitly we perform the colour algebra to yield the expression,
\ba
\lefteqn{{\cal{SLC}}\Big(\sum_{(i,j)}\bs{J}_{2}^{(1)}(i,j)\ 2{\rm{Re}}\la{\cal{A}}_{4}^{0}|\bt_{i}\cdot\bt_{j}|{\cal{A}}_{4}^{1}\ra\Big)=N^2(N^2-1)\frac{12}{2!}\sum_{(i,j)\in P(3,4)}}\nn\\
2{\rm{Re}}&\Big\{&\Big(\bs{J}_{2}^{(1)}(\hb{1},i)+\bs{J}_{2}^{(1)}(\hb{2},j)-\bs{J}_{2}^{(1)}(\hb{1},j)-\bs{J}_{2}^{(1)}(\hb{2},i)\Big){\cal{A}}_{4}^{0\dagger}(\hb{1},\hb{2},i,j){\cal{A}}_{4,1}^{1}(\hb{1},\hb{2},j,i)\nn\\
&+&\Big(\bs{J}_{2}^{(1)}(\hb{1},i)+\bs{J}_{2}^{(1)}(\hb{2},j)-\bs{J}_{2}^{(1)}(\hb{1},\hb{2})-\bs{J}_{2}^{(1)}(i,j)\Big){\cal{A}}_{4}^{0\dagger}(\hb{1},\hb{2},i,j){\cal{A}}_{4,1}^{1}(\hb{1},i,\hb{2},j)\nn\\
&+&\Big(\bs{J}_{2}^{(1)}(\hb{1},\hb{2})+\bs{J}_{2}^{(1)}(i,j)-\bs{J}_{2}^{(1)}(\hb{1},i)-\bs{J}_{2}^{(1)}(\hb{2},j)\Big){\cal{A}}_{4}^{0\dagger}(\hb{1},i,\hb{2},j){\cal{A}}_{4,1}^{1}(\hb{1},\hb{2},i,j)\Big\}.\nn\\\label{eq:m42poles1}
\ea
Once again, the mass factorization kernels used to define the integrated dipoles cancel.

The piece of the double virtual subtraction term proportional to the one-loop four gluon amplitudes is obtained by the analytic integration of $\dsigma_{NNLO}^{T,b_{1}}$,
\ba
\dsigma_{NNLO}^{U,a}&=&-\int_{1}\dsigma_{NNLO}^{T,b_{1}}\nn\\
&=&-{\cal N}_{LO}\left(\frac{\alpha_s}{2\pi}\right)^2\bar{C}(\epsilon)^2\int \frac{{\rm d}z_1}{z_1}\frac{{\rm d}z_2}{z_2}{\rm d}\Phi_{3}(p_3,p_4;\bar{p}_1,\bar{p}_2)\;J_{2}^{(2)}(p_{3},p_{4})\ \frac{24}{2!}\sum_{(i,j)\in P(3,4)}\nn\\
\times2{\rm{Re}}&\Big\{&\Big(\frac{1}{2}\calF(s_{\b{1}i})+\frac{1}{2}\calF(s_{\b{2}j})-\frac{1}{2}\calF(s_{\b{1}j})-\frac{1}{2}\calF(s_{\b{2}i})\Big){\cal{A}}_{4}^{0\dagger}(\hb{1},\hb{2},i,j){\cal{A}}_{4,1}^{1}(\hb{1},\hb{2},j,i)\nn\\
&+&\Big(\frac{1}{2}\calF(s_{\b{1}i})+\frac{1}{2}\calF(s_{\b{2}j})-\calF(s_{\b{1}\b{2}})-\frac{1}{3}\calF(s_{ij})\Big){\cal{A}}_{4}^{0\dagger}(\hb{1},\hb{2},i,j){\cal{A}}_{4,1}^{1}(\hb{1},i,\hb{2},j)\nn\\
&+&\Big(\calF(s_{\b{1}\b{2}})+\frac{1}{3}\calF(s_{ij})-\frac{1}{2}\calF(s_{\b{1}i})-\frac{1}{2}\calF(s_{\b{2}j})\Big){\cal{A}}_{4}^{0\dagger}(\hb{1},i,\hb{2},j){\cal{A}}_{4,1}^{1}(\hb{1},\hb{2},i,j)\Big\}.\nn\\\label{eq:dsigua}
\ea
The poles of Eq.~\eqref{eq:dsigua} match those of Eq.~\eqref{eq:m42poles1}.

\subsection*{Double operator insertions into tree-level sandwiches}

The second subset of poles contained in the two-loop interferences is written in terms of double charge operator insertions carrying poles given by convolutions of integrated dipoles,
\ba
\sum_{(i,j)}\sum_{(k,l)}\big[\bs{J}_{2}^{(1)}(i,j)\otimes\bs{J}_{2}^{(1)}(k,l)\big]\ \la{\cal{A}}_{4}^{0}|(\bt_{i}\cdot\bt_{j})(\bt_{k}\cdot\bt_{l})|{\cal{A}}_{4}^{0}\ra.
\ea
Evaluating the colour sums explicitly and keeping only the sub-leading colour contribution yields,
\ba
\lefteqn{\hspace{-3cm}{\cal{SLC}}\Big(\sum_{(i,j)}\sum_{(k,l)}\big[\bs{J}_{2}^{(1)}(i,j)\otimes\bs{J}_{2}^{(1)}(k,l)\big]\ \la{\cal{A}}_{4}^{0}|(\bt_{i}\cdot\bt_{j})(\bt_{k}\cdot\bt_{l})|{\cal{A}}_{4}^{0}\ra\Big)=N^2(N^2-1)}\nn\\
\frac{12}{2!}\sum_{(i,j)\in P(3,4)}\frac{1}{2}&\bigg[&\Big(\bs{J}_{2}^{(1)}(\hb{1},i)+\bs{J}_{2}^{(1)}(\hb{2},j)-\bs{J}_{2}^{(1)}(\hb{1},j)-\bs{J}_{2}^{(1)}(\hb{2},i)\Big)\nn\\
&&\otimes\Big(\bs{J}_{2}^{(1)}(\hb{1},i)+\bs{J}_{2}^{(1)}(\hb{2},j)-\bs{J}_{2}^{(1)}(\hb{1},\hb{2})-\bs{J}_{2}^{(1)}(i,j)\Big)A_{4}^{0}(\hb{1},\hb{2},i,j)\nn\\
&+&\frac{1}{2}\Big(\bs{J}_{2}^{(1)}(\hb{1},\hb{2})+\bs{J}_{2}^{(1)}(i,j)-\bs{J}_{2}^{(1)}(\hb{1},j)-\bs{J}_{2}^{(1)}(\hb{2},i)\Big)\nn\\
&&\otimes\Big(\bs{J}_{2}^{(1)}(\hb{1},\hb{2})+\bs{J}_{2}^{(1)}(i,j)-\bs{J}_{2}^{(1)}(\hb{1},i)-\bs{J}_{2}^{(1)}(\hb{2},j)\Big)A_{4}^{0}(\hb{1},i,\hb{2},j)\bigg].\label{eq:JxJpoles}
\ea
The relevant piece of the double virtual subtraction term is constructed from the analytic integration of the real-virtual subtraction terms, $\dsigma_{NNLO}^{T,b_{2}}$ and $\dsigma_{NNLO}^{T,c_{2}}$ and the double real subtraction term $\dsigma_{NNLO}^{S,d}$,
\ba
\dsigma_{NNLO}^{U,b}&=&-\int_{1}\Big[\dsigma_{NNLO}^{T,b_{2}}+\dsigma_{NNLO}^{T,c_{2}}\Big]-\int_{2}\dsigma_{NNLO}^{S,d}\nn\\
&=&-{\cal N}_{LO}\left(\frac{\alpha_s}{2\pi}\right)^2\bar{C}(\epsilon)^2
\int \frac{{\rm d}z_1}{z_1}\frac{{\rm d}z_2}{z_2}~{\rm d}\Phi_{3}(p_3,p_4;\bar{p}_1,\bar{p}_2)\;J_{2}^{(2)}(p_{3},p_{4})\frac{24}{2!}\sum_{(i,j)\in P(3,4)}\nn\\
\frac{1}{2}&\Big\{&\Big(\frac{1}{2}\calF(s_{\b{1}i})+\frac{1}{2}\calF(s_{\b{2}j})-\frac{1}{2}\calF(s_{\b{1}j})-\frac{1}{2}\calF(s_{\b{2}i})\Big)\nn\\
&&\otimes\Big(\frac{1}{2}\calF(s_{\b{1}i})+\frac{1}{2}\calF(s_{\b{2}j})-\calF(s_{\b{1}\b{2}})-\frac{1}{3}\calF(s_{ij})\Big)A_{4}^{0}(\hb{1},\hb{2},i,j)\nn\\
&+&\frac{1}{2}\Big(\calF(s_{\b{1}\b{2}})+\frac{1}{3}\calF(s_{ij})-\frac{1}{2}\calF(s_{\b{1}j})-\frac{1}{2}\calF(s_{\b{2}i})\Big)\nn\\
&&\otimes\Big(\calF(s_{\b{1}\b{2}})+\frac{1}{3}\calF(s_{ij})-\frac{1}{2}\calF(s_{\b{1}i})-\frac{1}{2}\calF(s_{\b{2}j})\Big)A_{4}^{0}(\hb{1},i,\hb{2},j)\Big\}.\label{eq:dsub}
\ea
It can be easily seen that the poles of Eq.~\eqref{eq:dsub} match those of Eq.~\eqref{eq:JxJpoles}.

At this point, we have shown that all explicit poles of the two-loop matrix elements cancel against $\dsigma_{NNLO}^{U,a}$ and $\dsigma_{NNLO}^{U,b}$ and there are no further contributions from the analytic integration from either the double real or real-virtual subtraction terms,
\ba
\Poles\bigg(\dsigma_{NNLO}^{VV}-\dsigma_{NNLO}^{U}\bigg)&=&0.
\ea

\section{Numerical evaluation of the differential cross section}

In Secs.~\ref{sec:RR},~\ref{sec:RV} and~\ref{sec:VV} the double real, real-virtual and double virtual subtraction terms were constructed and, where appropriate, the explicit pole cancellation against one and two-loop matrix elements at sub-leading colour was carried out. The remaining task is to numerically integrate each of these partonic channels over the appropriate phase space to obtain the physical cross section. 

Our numerical studies for proton-proton collisions at centre-of-mass energy $\sqrt{s}=8$ TeV concern the single jet inclusive cross section (where every identified jet in an event that passes the
selection cuts contributes, such that a single event potentially enters the distributions multiple times) and the two-jet exclusive cross section (where events with exactly two identified jets contribute). 
We use in our default setup the anti-$k_t$ jet algorithm~\cite{Cacciari:2008gp} with resolution parameter $R=0.7$ to reconstruct the final state jets where jets are accepted at central rapidity $|y| < 4.4$, and ordered in transverse
momentum. An event is retained if the leading jet has $p_{T1} > 80$ GeV. For the dijet invariant mass distribution, a second jet must be observed with $p_{T2} > 60$ GeV.

All calculations are carried out with the MSTW08NNLO gluon 
distribution function~\cite{Martin:2009bu}, including the evaluation of the 
LO and NLO contributions.\footnote{Note that the evolution of the gluon distribution within the PDF set together with the value of $\alpha_s$ intrinsically includes contributions from the light quarks.  The NNLO calculation presented here is ``gluons-only" in the sense that only gluonic matrix elements are involved.} This choice of parameters
allows us to quantify the size of the genuine NNLO contributions to the  
parton-level subprocess. As default value, we
set $\mu$ equal to the transverse momentum of the leading jet so that $\mu = p_{T1}$. 

\begin{figure}[t]
\centering
\includegraphics[width=0.71\textwidth]{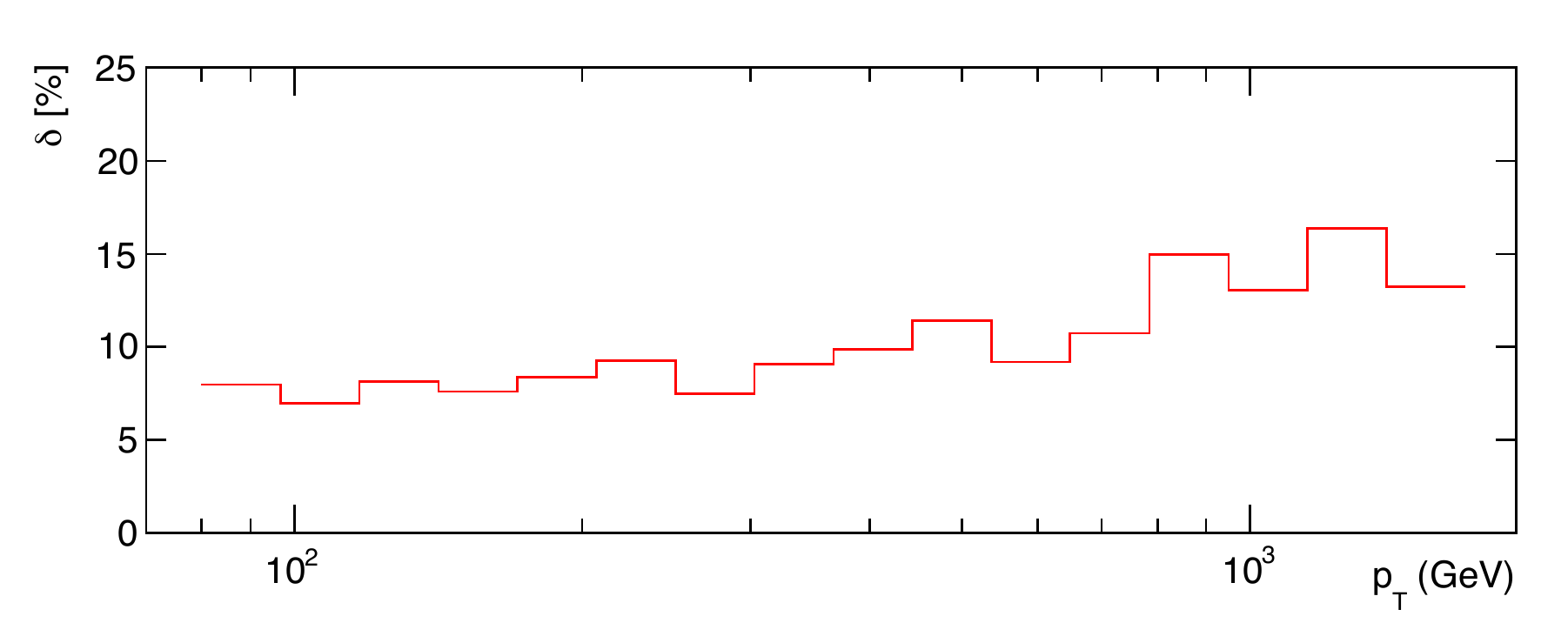}\\\vspace{-0.2cm}
\caption{The percentage contribution of the sub-leading colour to full colour NNLO correction, $\delta$, for the single jet inclusive transverse energy distribution as a function of $p_{T}$.}
  \label{fig:subratio}
\end{figure}

The cross section can be written as,
\begin{equation}
{\rm{d}}\sigma = \alpha_s^2 A +\alpha_s^3 B +\alpha_s^4 C,
\end{equation}
where the coefficients $A$, $B$ and $C$ depend on the PDF, the scale choice and the observable.
The NNLO coefficient $C$ can be further subdivided into leading and sub-leading colour contributions,
\begin{equation}
C = C^{LC} + C^{SLC}.
\end{equation}
To quantify the size of the sub-leading colour NNLO corrections, Fig.~\ref{fig:subratio} shows the ratio,
$$\delta = \frac{C^{SLC}}{C}$$
as a percentage for the single jet inclusive transverse energy distribution.  We see that $\delta$ is roughly 10\% as expected from naive power counting of colours ($1/N^2$), but exhibits a $p_{T}$ dependence, rising from
8\% at low $p_{T}$ to 15\% at high $p_{T}$.

\begin{figure}[t]
\centering
\includegraphics[width=0.71\textwidth]{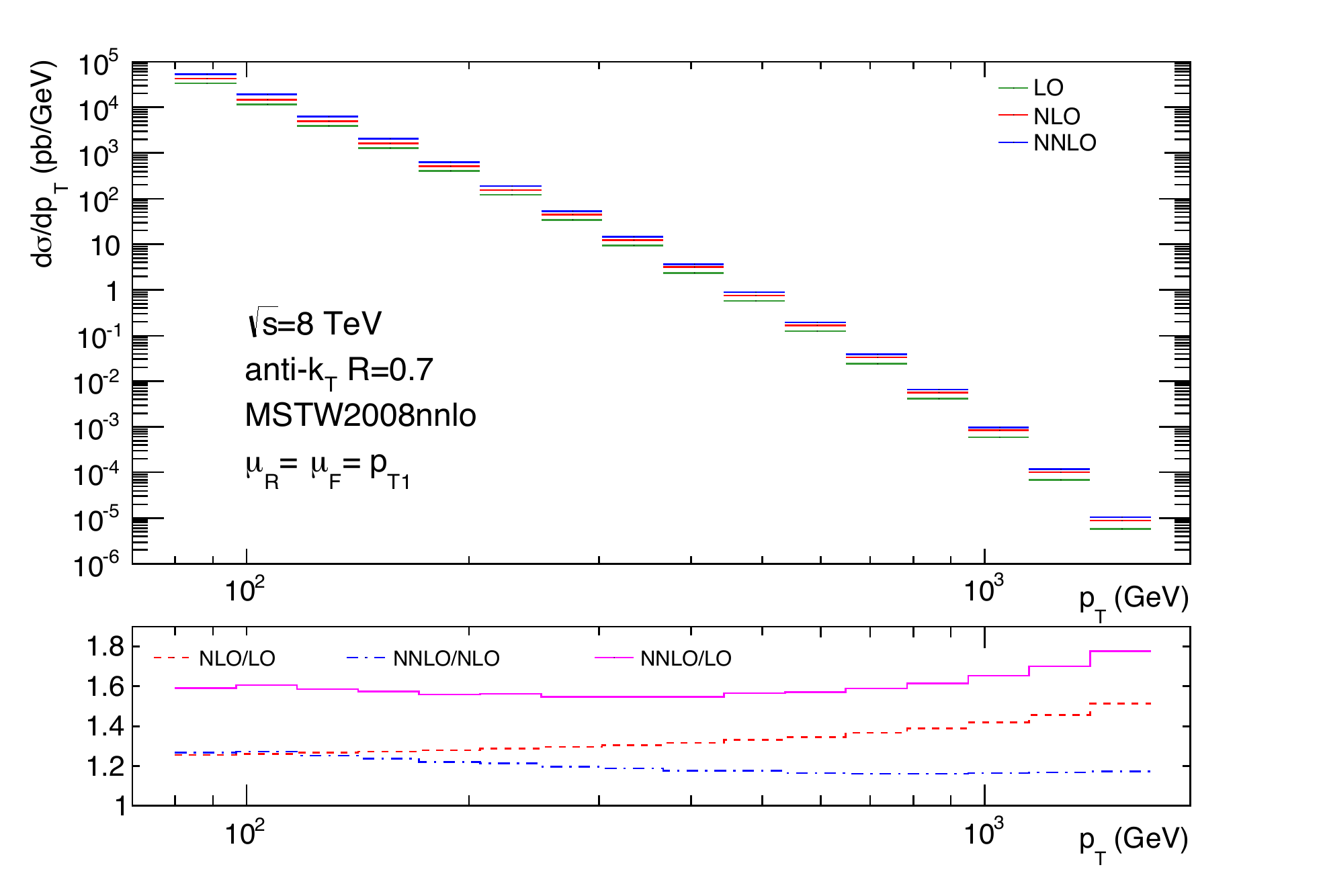}
\caption{Inclusive jet transverse energy distribution, $d\sigma/dp_T$, for jets constructed with the anti-$k_T$ algorithm with $R=0.7$ and with $p_T > 80$~GeV, $|y| < 4.4$ and $\sqrt{s} = 8$~TeV at NNLO (blue), NLO (red) and LO (dark-green). The lower panel shows the ratios of NNLO, NLO and LO cross sections.}
  \label{fig:dsdet}
\end{figure}

In Fig.~\ref{fig:dsdet} we present the inclusive jet cross section for the
anti-$k_T$ algorithm with $R=0.7$ and with $p_T > 80$~GeV, $|y| < 4.4$ as a
function of the jet $p_{T}$ at LO, NLO and NNLO, for the central scale choice
$\mu = p_{T1}$ retaining the full dependence of the number of colours. The NNLO/NLO $k$-factor shows the ratio of the NNLO
and NLO cross sections  in each bin.
For this scale choice we see that the NNLO/NLO $k$-factor across the $p_{T}$ range corresponds to a 16-26\% increase compared to the NLO
cross section. 

\begin{figure}[h!]
\centering
\begin{minipage}[b]{0.45\linewidth}
  \includegraphics[width=0.9\textwidth]{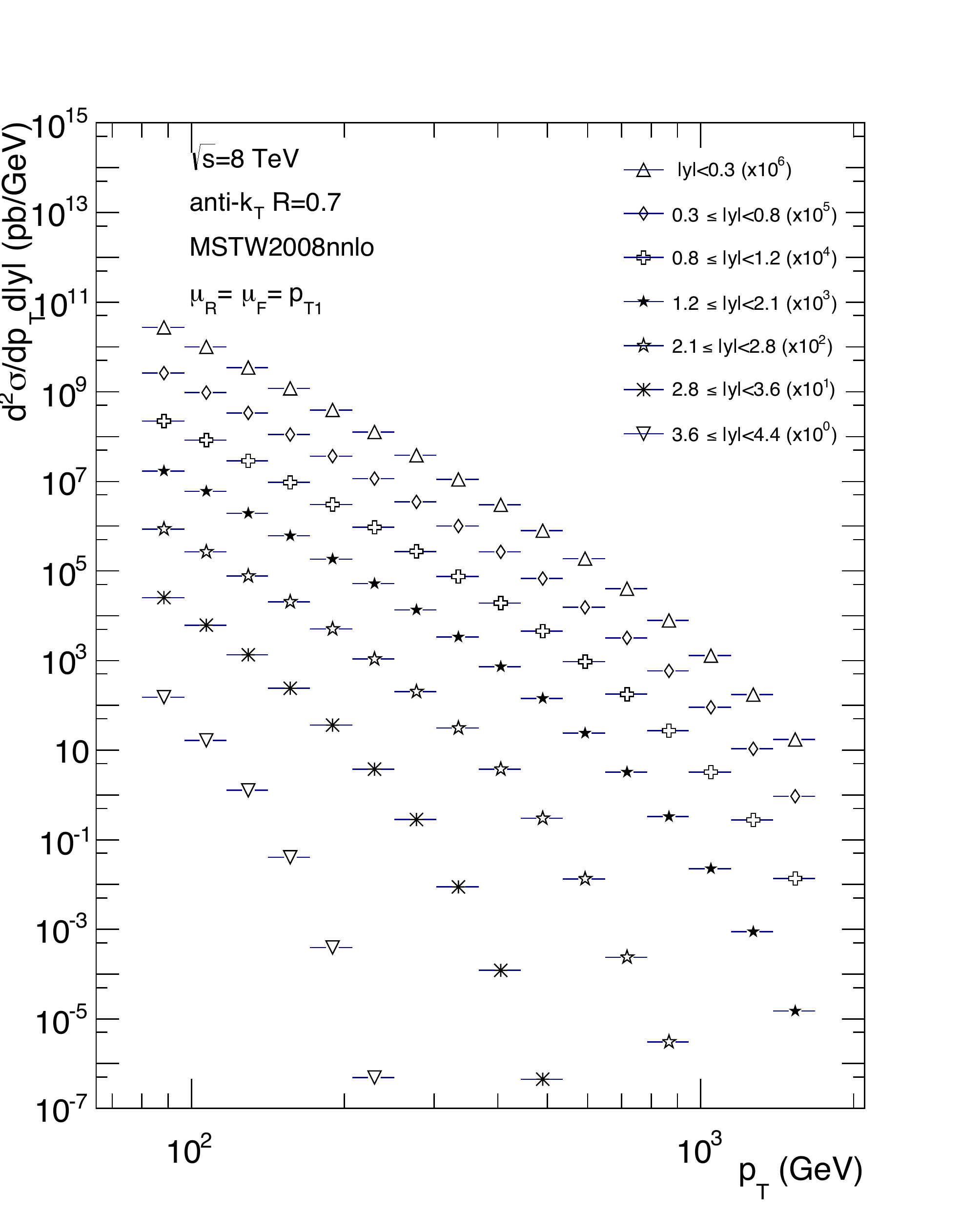}
\end{minipage}
\quad
\begin{minipage}[b]{0.45\linewidth}
  \includegraphics[width=1.1\textwidth]{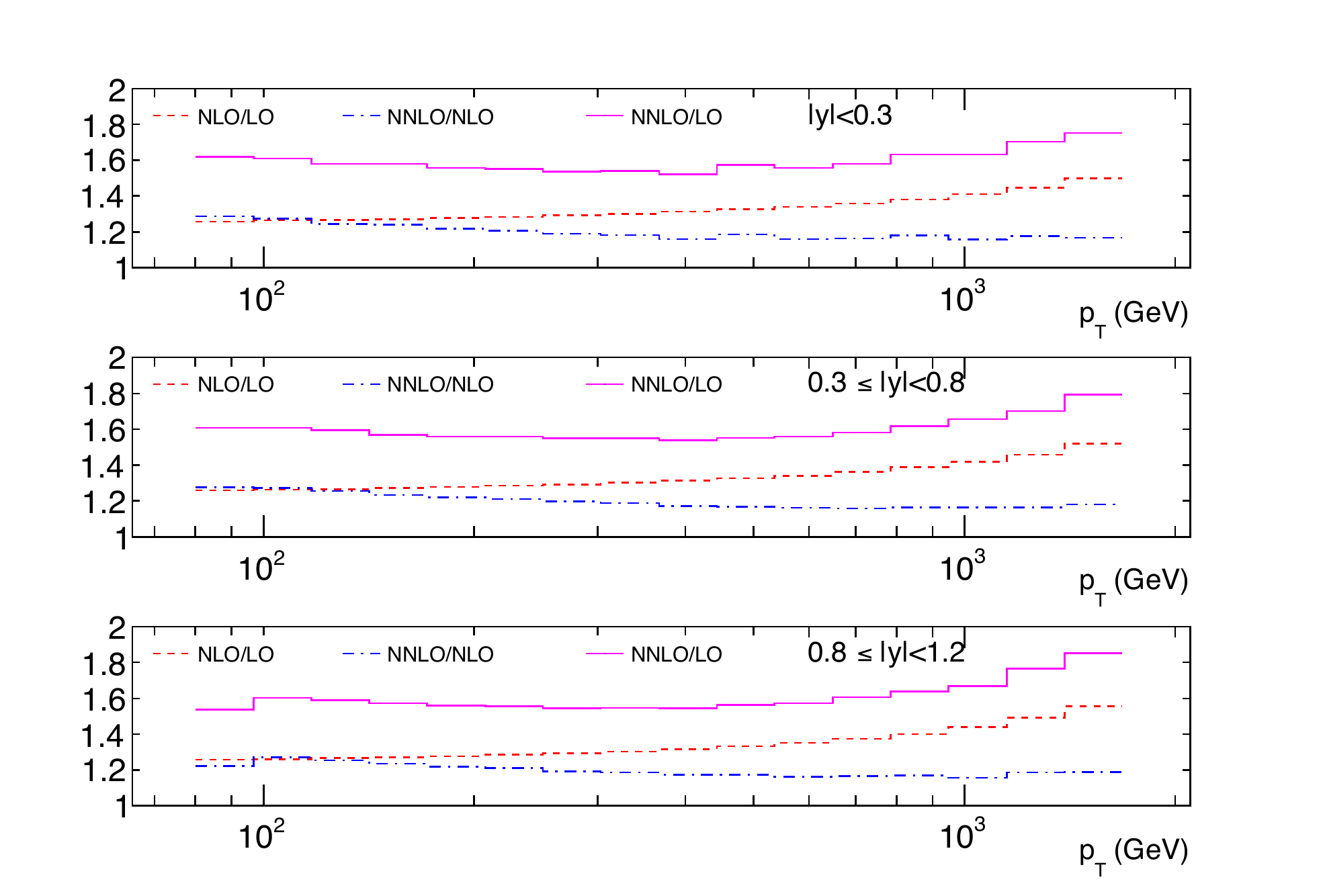}\\
\end{minipage}
  \caption{The left panel shows the doubly differential inclusive jet transverse energy distribution, $d^2\sigma/dp_T d|y|$, at $\sqrt{s} = 8$~TeV for the anti-$k_T$ algorithm with $R=0.7$ and for $p_T > 80$~GeV and various $|y|$ slices at NNLO. 
  The right panel shows the ratios of NNLO, NLO and LO cross sections for three rapidity slices: $|y | < 0.3$, $0.3 < |y| < 0.8$ and $0.8 < |y| < 1.2$.}
  \label{fig:d2sdetslice}
\end{figure}
\begin{figure}[h!]
\centering
\begin{minipage}[b]{0.45\linewidth}
  \includegraphics[width=0.9\textwidth]{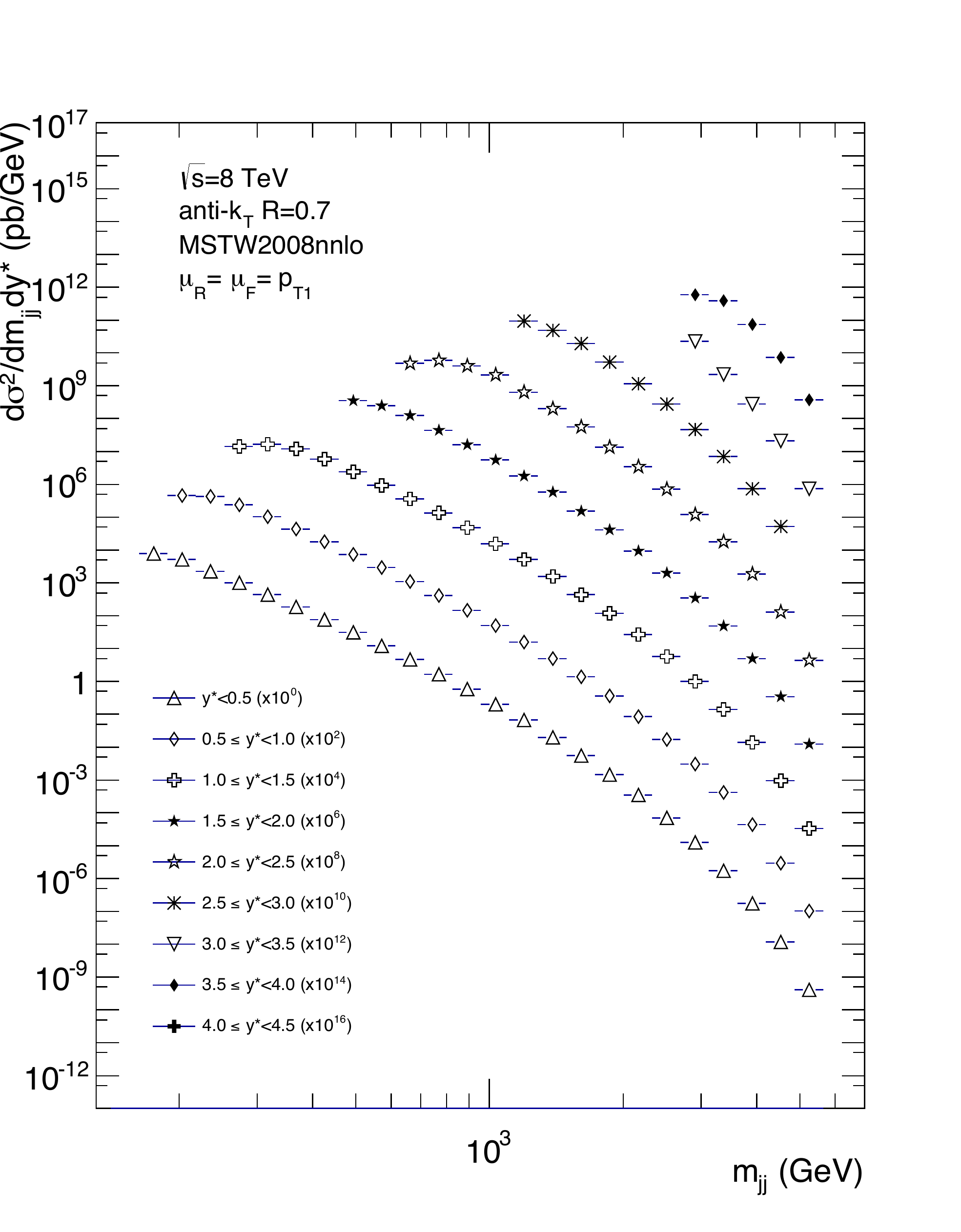}
\end{minipage}
\quad
\begin{minipage}[b]{0.45\linewidth}
  \includegraphics[width=1.1\textwidth]{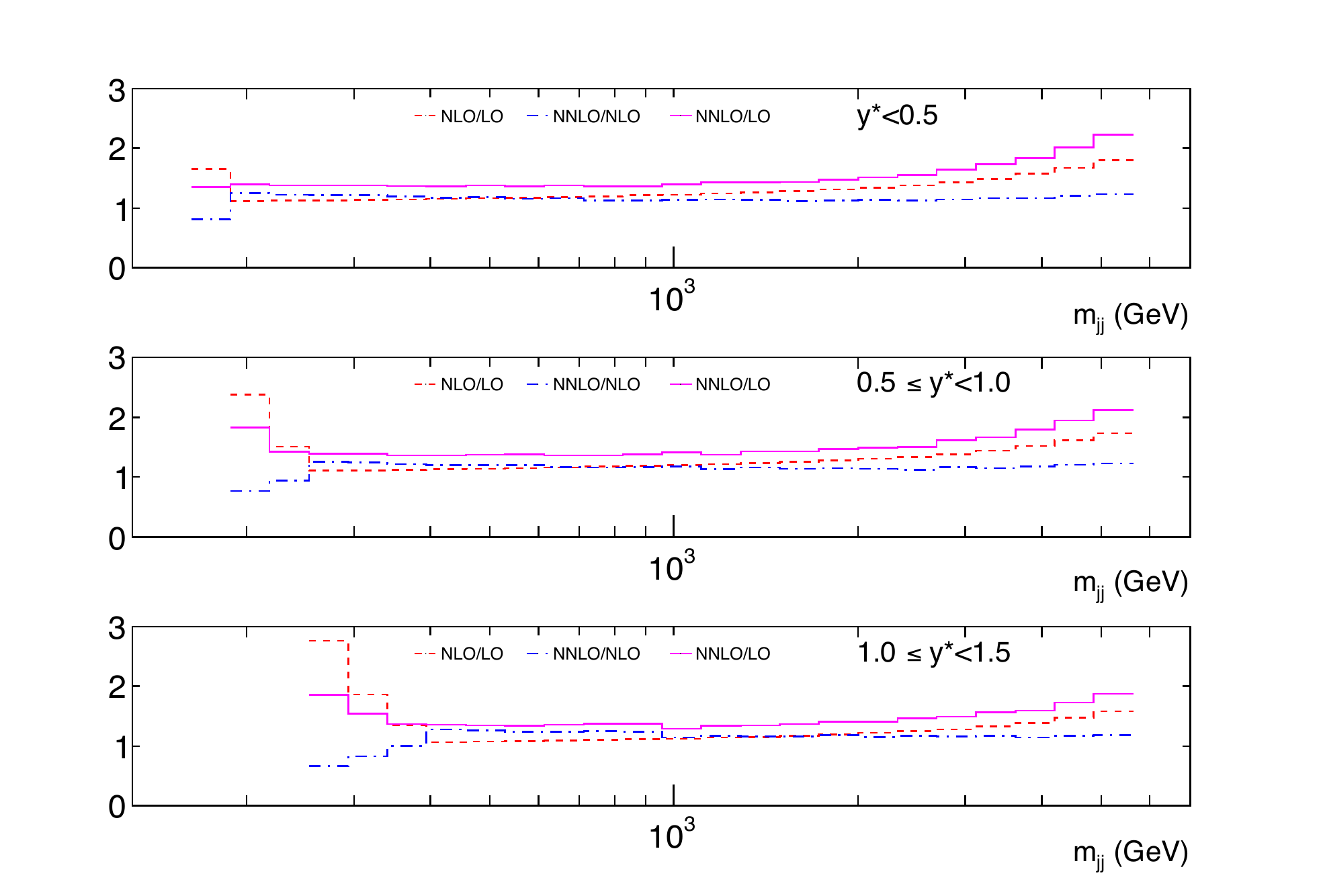}\\
\end{minipage}
  \caption{The left panel shows the doubly differential exclusive dijet invariant mass distribution, $d^2\sigma/dm_{jj} dy^*$, at $\sqrt{s} = 8$~TeV for the anti-$k_T$ algorithm with $R=0.7$ and for $p_{T1} > 80$~GeV, 
  $p_{T2}>60$~GeV and various $y^*=|y_{1}-y_{2}|/2$ slices at NNLO. The right panel shows the ratios of NNLO, NLO and LO cross sections for three rapidity slices: $y^* < 0.5$, $0.5 < y^* < 1.0$ and $1.0 < y^* < 1.5$.}
  \label{fig:d2sdmjjslice}
\end{figure}

In Fig.~\ref{fig:d2sdetslice} we present the inclusive jet cross section in double differential form. The inclusive jet cross section
is computed in jet $p_{T}$ and rapidity bins over the range 0.0-4.4 covering central and forward jets. To quantify the impact of the NNLO correction we present the double differential $k$-factors containing 
ratios of NNLO, NLO and LO cross sections in the same figure. We observe that the NNLO correction increases the cross section between 26\% at low $p_{T}$ to 14\% at high $p_{T}$ with respect to the 
NLO calculation. This behaviour is similar for each of the three rapidity slices presented.

As a final observable, we computed the exclusive dijet cross section at NNLO. For this cross section we require two jets in the final state from which we reconstruct the invariant mass of the dijet system and compute
the double differential dijet cross section in bins of invariant mass $m_{jj}$ and $y^*=|y_{1}-y_{2}|/2$ slices over the range 0.0-4.5. The results at NNLO are presented in Fig.~\ref{fig:d2sdmjjslice}. 
The exclusive dijet events are a subset of the inclusive jet events and we observe that the NNLO/NLO $k$-factor is approximately flat across the $m_{jj}$ range corresponding to a 16-21\% 
increase when compared to the NLO cross section.

\section{Summary}

In this paper we have computed the full colour contributions to jet production from gluon scattering at NNLO. Previous work~\cite{Glover:2010im,GehrmannDeRidder:2011aa,GehrmannDeRidder:2012dg,GehrmannDeRidder:2013mf} focussed on the leading colour contribution. The new element is the inclusion of the sub-leading colour effects which contribute first at NNLO. Unlike at leading colour, the double real and real-virtual contributions cannot be written in terms of squared partial amplitudes, but appear as interferences of different colour ordered amplitudes.

To isolate the soft singularities we used the antenna subtraction technique which required no significant alterations or new ingredients in order to deal with the incoherent interferences of partial amplitudes.  We found that the single and double unresolved limits of the double real matrix element at sub-leading colour could be fully described using just three-parton tree-level antennae and soft factors, without the need for four-parton antenna functions.  Similarly, the single unresolved limits of the real-virtual matrix element did not require the one-loop three-parton antenna and could be described with
only tree-level three-parton antennae to remove all explicit and implicit singularities. In the process, we found a very compact form for the real-virtual matrix element which we believe to be a new addition to the literature. The double virtual subtraction term, generated by integrating the remaining double real and real-virtual subtraction terms, also involves incoherent interferences of four-parton one-loop and tree-level amplitudes. We showed that it analytically cancels the explicit poles present in the formula for the two-loop matrix elements \cite{Glover:2001af,Glover:2001rd}.

With the double real, real-virtual and double virtual subtraction terms in place, the matrix elements are free from explicit poles in $\e$ and finite in all unresolved regions of phase space and so can be numerically integrated in four dimensions to produce finite corrections to the physical distributions. This work provides the first quantitative estimate for the size of sub-leading colour contributions to jet production relative to the leading-colour approximation.  The corrections are found to be in line with prior expectations, providing approximately a 10\% correction to the NNLO leading colour contribution. This completes the study of jet production at NNLO in the all-gluon approximation; future work will move beyond this approximation and include scattering processes involving light quarks.

\acknowledgments 

We thank Thomas Gehrmann for useful discussions, constructive comments and reading of the manuscript. 
This research was supported by the Swiss National Science Foundation
(SNF) under contract PP00P2-139192, in part by the European Commission through the `LHCPhenoNet' Initial Training Network
PITN-GA-2010-264564, in part by the UK Science and Technology Facilities Council through grant ST/G000905/1 and in part by the National Science Foundation under Grant No. PHY11-25915.
AG would like to thank the Kavli Institute for Theoretical Physics, Santa Barbara, for its kind hospitality, 
where part of this work was carried out. EWNG gratefully acknowledges the support of the Wolfson Foundation and the Royal Society and thanks the Institute for Theoretical Physics at the ETH and the Pauli Center for Theoretical Studies for their kind hospitality during the completion of this work.

\bibliographystyle{JHEP}
\bibliography{ref}

\end{document}